# The reach of a verification tool decides its value: A controlled study of verification surface, artifact quality, and cost in AI coding agents

**Achint Mehta**

**ABSTRACT** Modern artificial-intelligence coding agents can be equipped with tools for checking their own work e.g. a linter, a boot probe, a shell, a screenshot tool. We call this set the agent's verification surface. This study asks whether increasing only that surface, with everything else held fixed, produces a matching growth in the quality of the software the agent ships. We built a minimal coding agent whose tool list is the single controlled variable and used it to implement 1,116 web applications across six models and eight tool configurations. A condition-blind human graded every application against a frozen rubric, and automatic probes stress-tested the API-observable behaviors. Verification's cheapest benefit arrives first, which is to make sure that the application comes up. Without any tools, about one build in seven fails to launch at all and a single boot probe removes nearly all of these failures at roughly 35 percent of a full shell's token cost, while the full shell multiplies the no-tools cost by 2.35. Screenshots help most where mistakes are visible (e.g. element placement, interaction), though even there the gain over a shell is modest and does not survive correction for multiple statistical comparisons. In cases where failures can only be measured rather than seen, such as keeping scrolling smooth over a 100,000-row list, screenshots add nothing. A verification tool improves the output artifact only where its reach covers the way the application actually fails.

**INDEX TERMS** Artificial intelligence, benchmark testing, large language models, program testing, software agents, software quality, software reliability

## I. INTRODUCTION

### A. Goal

This study asks how the verification surface of an artificial-intelligence coding agent, the set of tools through which the agent can observe and check its own work, affects the quality of the software the agent actually ships i.e. its output files, as opposed to the transcript of what it said while working.

To change only the verification surface while holding everything else constant, we built a custom, minimal coding agent. Because the agent is ours, the system prompt is fixed, and the only thing that changes between conditions is the list of verification tools the agent is handed. The agent ran with one of five core tool sets: none at all, a linter only, a boot probe only, a full shell, or the shell plus screenshots (three further targeted configurations are described in Section II). A linter reads code and flags likely mistakes without running it. The boot probe is a single check that installs dependencies, starts the application's server, and reports whether it listens. The full shell is an unrestricted command line that can run anything.

Six models, claude-4.6-sonnet, claude-4.6-opus, claude-4.8-opus, gpt-5.5, gemini-3.1-pro, and grok-4.3, each implemented seven web applications under these configurations, four to five times per combination. We call these repeats runs replicates and each replicate was an independent run which received no feedback from any other run. In total, 1,116 applications were built. A human grader then scored every application in shuffled order against a fixed rubric, and an automatic probing tool checked the subset of behaviors observable through the API.

### B. Contributions

#### 1) A controlled dose-response measurement

Similar to medicine's dose-response studies, each rung of our verification tools ladder adds one capability and we measure what the increment buys. With our custom coding agent, we controlled every configuration and provided tools to the model via this coding agent in an ordered ladder, with each configuration increasing the verification capability over the last. These capabilities are: none, linter only, boot probe only, full shell, and shell plus screenshots. Because we controlled everything else, only a single parameter changed between conditions. This lets us tie the outcomes to the tools rather than other uncontrollable environment parameters.

*2) The boot probe result*

The boot probe tool (install dependencies and see if the server starts or not) recovered nearly all the launch failures observed with the "no tools" configuration. This single-tool configuration gave the highest ratio of artifact quality to token usage (tokens are the unit in which model usage is billed, so token counts are our measure of cost in this study).

*3) A more nuanced look into the question "does sight help"*

In our study we measure where the value of sight (the screenshot tool) starts and stops. The likeliest gains from sight over a shell (which allows the model to run any arbitrary command) alone appear where the failures involve element placement or interaction. Though this visual advantage is modest and does not survive correction for multiple comparisons, we treat this as suggestive rather than established. The gains are smaller for static page-layout issues, which a model can often get right from the code alone, without seeing the page. Whether a section is present and roughly arranged is visible in the markup, while whether elements collide or misalign at real sizes is not. It also seems to provide minimal value in detecting issues that can be measured but not seen through code such as drag-and-drop that flickers, or scrolling that is not smooth.

### C. Prior Work

Prior work shows that the scaffolding around an artificial-intelligence coding agent, its prompts, tools, and control loop, strongly affects output quality [1]. Those studies compare whole systems against whole systems but this study changes a single part of the scaffold and measures what that part buys. Our tasks were chosen to make the measurement possible with some applications are API-heavy while others are visually demanding.

The design choices for this study and our test harness were also motivated by the limitations of existing studies. SWE-agent [2] showed that interface design alone can roughly triple task resolution. Agentless [3] showed a simple pipeline competitive with full agents at a tenth of the cost. These existing works also compare complete systems by counting how many benchmark problems each one solves, in the pass-rate tradition running from HumanEval [4], SWE-bench [5], and WebGen-Bench [6]. In these studies, a problem counts as solved only if the project's own tests pass afterward, and each attempt earns one verdict, pass or fail. In our study, we hold most of the system constant while changing one aspect of the system, and our outcome is quality rather than a simple pass/fail rate.

CodeAgent [7] and the self-correction literature [8] propose systems in which running code and reading the results improve outcomes, with Reflexion [9] and Self-Debugging [10] the canonical demonstrations that a model can improve its own output from execution feedback. These papers propose a system and present evidence that it works. Ours holds the system fixed and measures which part of the verification surface buys which kind of quality.

In our own previous study [11], we examined a related question observationally with 90 implementations of the same application (a real-time retrospective board) across model tiers, two agent harnesses, reasoning-effort levels, an interactive browser-testing tool, and design-oriented prompts. Its headline findings were that reasoning effort, not tool access, bought first-try reliability, and that the browser-testing tool raised cost by 42 to 68 percent without improving functional score.

That study did not conclude that tools are useless. Its claim compared tools usage with reasoning effort in a much more general sense. Raising effort improved reliability across every kind of failure at once, whereas a tool only helps with the failures it can actually observe, so in that study no tool configuration we tried matched what a step up in reasoning bought. The failures there were mostly environment failures, with Docker deployment failing first try in 44 percent of runs and the local development setup in another 17 percent, and a tool that exercises the finished page cannot see either. That paper noted that a tool aimed at those faults, such as a container build or smoke test, could have caught them.

The present study takes up exactly that unanswered question. Instead of asking whether tools can beat a better model, it focuses on the tools themselves, and which part of the verification surface buys which kind of quality, and at what cost. The earlier study varied model capability and found it dominant. This study holds capability still, running every model in every condition with byte-for-byte identical inputs, so that what remains is the effect of the tools themselves.

The two studies are otherwise unrelated in practice. This one uses an agent we built rather than existing harnesses, seven applications rather than one, 1,116 hands-off runs rather than ninety runs with a human in the loop, and hypotheses fixed in a written analysis plan before any statistic was computed. No data, runs, or analyses are shared between them.

Another set of studies [12], [13], [14], [15], [16], [17] establishes that visual and automated feedback improve frontend generation (the frontend is the part of an application that runs in the browser and draws the page), from static screenshot-to-code translation [16] to iterative visual refinement [12], [15] and feedback-trained generation [17], and we are not trying to replicate that. Our study differs in various aspects from them. First, those studies evaluate visually defined tasks only, so they cannot ask when sight stops paying. Our selection of applications that the models were asked to implement, deliberately crosses visually hard tasks with applications whose correctness is fully observable through the APIs, and we find the visual advantage shrinks toward zero on the latter. Second, their systems grant vision ability alongside execution, while our sight-without-shell condition separates the two abilities and suggests that the value can be derived from the sight ability itself. Third, their outcome is task success, while ours is a set of more granular criteria (startup survival, behavior under stress probes, human-judged interface correctness, cost, and variance), several of which could have opposite effect under the same conditions.

Coding with Eyes argues that GUI [13] (Graphical User Interface, another way of saying frontend) code requires interaction, not just rendering, to verify. One of our runs exhibited exactly this. One page rendered perfectly but ended up ignoring every click (Section V). This is direct confirmation, and it implies our passive screenshot tool gives a lower bound on what sight could deliver and there is scope for improvement even after provisioning the screenshot tools.

Finally, a related line of work evaluates agents inside interactive environments such as WebArena [18] and OSWorld [19], where success depends on what the agent can observe of a live browser or desktop. Those studies vary the task world while keeping the agent's abilities fixed while ours holds the tasks fixed and varies only what the agent can observe of its own work.

The rest of this paper is organized as follows. Section II describes the agent, the seven applications, the tool conditions, and the grading instruments. Section III states the six pre-specified hypotheses and the statistical plan for this study. Section IV reports the results, Section V documents some of the reasons for the applications failures, and Section VI examines the configuration integrity and model signatures. Section VII states the limitations, and Section VIII concludes the paper.

## II. Study Design

### A. The agent

For this study we created a minimal coding agent instead of using a commercial or open-source coding agent for three main reasons. First, our agent is fully under our control and we can change only the tool configuration while keeping everything else constant. Second, we could log every input and output, including tool calls (which, when, and how) and token usage. We could verify whether a particular tool was called by the model or not. Third, it allowed us to control the system prompt and the input prompt. Existing harnesses, including open platforms such as OpenHands [20], carry their own system prompts and control-loop instructions [1], which are difficult to hold constant across conditions and could contaminate the results. Our agent has one fixed base prompt, one fixed environment briefing that is identical across all runs, and the task text. Nothing else reaches the model.

Our agent loop is kept intentionally simple. The model receives the prompt and tool schemas (machine-readable descriptions of each tool that the model may call) and then calls the tools until it declares completion of the task by calling a special tool named *finish*. The model can receive at most one nudge. If the model stops calling tools without having declared completion, the agent sends it a single fixed reminder message asking it to continue or finish. The nudge is applied by a fixed rule identical across conditions and only prompts the model to declare completion. It does not add verification ability or change the artifact. In 29 of the 1,116 runs, the model went idle without declaring completion, and a single nudge from the agent prompted it, after which all declared completion. In terms of model breakdown gpt-5.5 has 20, claude-4.6-sonnet has 4, grok-4.3 has 3, and claude-4.8-opus has 2 implementations that required a single nudge to reach completion.

Basic file tools (read, write, edit, list) were available to the model in every condition. The conditions differed only in which verification tools were made available to the model on top of the basic file tools.

The agent requested no extended-reasoning mode from any model. There were instances where the gateway included a reasoning container in its response, but that container was empty. Across the 1,116 published traces the gateway returned 27,083 such containers, none of them carrying content. No reasoning was therefore available for the agent to carry from one turn to the next, and none was sent back. Models may still reason internally before answering, and where they do those tokens are included in the totals we report, but that reasoning was never exposed to the agent and so was never available to carry from one turn to the next. Each turn therefore proceeds from that message history alone.

### B. The applications

Each application was defined as a frozen written specification (the spec) with an explicit "Acceptance criteria" section. The specs did not mention tools, so the same spec was valid in every condition. Keeping tools schema and task spec separate also allowed models to pick the tools based on their reasoning ability rather than being influenced by user instructions.

Each spec was chosen to target a specific set of behaviors and failures, so that a particular kind of verification could plausibly prevent it.

The following specs were written as part of this study:

1. *message-board*: A minimal real-time message board which requires the user should be able to post a message, see the history, and see new messages appear live in other open windows (via server-sent events, a standard browser mechanism for pushing updates). This spec defines the calibration floor of the study and is deliberately easy enough that most of the implementations should succeed at this.
2. *kanban-board*: A shared task board with columns and draggable cards, where card ordering uses fractional positions that must be kept consistent under repeated reordering, and where two open windows must converge to the same board state. Ordering under stress and convergence are exactly the properties an API can be made to hammer and this is one of the applications for which the API probes were written, while the feel of drag-and-drop exists only in a browser and was verified by the human grading. This spec targets ordering logic under stress, real-time convergence, and the feel of drag-and-drop, which only exists in a browser and would be hard to verify from APIs.

3. *seat-booking*: A ticket booking application that presents the user with a visual theater-style seat map to select the seats, with temporary holds that expire, all-or-nothing multi-seat acquisition, and a hard guarantee that two users trying to book the same seat cannot both win. This spec targets concurrency correctness, which cannot be settled by just reading the source. Locking or transactional handling is itself visible in the code, but whether it addresses every scenario through which two simultaneous requests can both succeed in booking the same seat depends on timing which the code does not show. The only way to know is to run the application and make two conflicting requests race against each other. That racing procedure is precisely what the automated API probes automate, condition-blind, against every implementation.
4. *calendar-week-view*: A week calendar whose difficulty is almost entirely visual. The overlapping events must share column width exactly while they overlap, reclaim full width when they do not, and sit at pixel positions proportional to their times. The API is trivial, and a wrong implementation returns perfectly correct data while drawing a wrong picture. This spec targets failures that only sight can catch.
5. *metrics-dashboard*: A statistics dashboard graded at three screen widths (phone, tablet, desktop), with a chart, a dark mode that must restyle everything dynamically, and deliberately awkward dataset (an overlong label, a seven-digit number) that requires careful implementation to prevent sloppy layout. This spec targets a second, different family of sight-only failures, so that we do not have to rely on one app for visual conclusions.
6. *kanban-labels-brownfield*: This is the sole modification task (brownfield is the industry term for changing existing code, as opposed to greenfield work that starts fresh). The agent receives an existing, working kanban application that it did not write, plus a change request (add colored labels and filtering) with the constraint that nothing existing may break. This spec targets regression safety when changing unfamiliar code, which is one of the most common real-world software tasks.
7. *log-explorer-perf*: A log viewer with over 100,000 rows, strict latency requirements, a windowed API with page size limited to 200 rows per response, and a virtualized list (the page keeps only a bounded number of row elements in the document no matter where the user scrolls). A simple but flawed implementation could return identical-looking data and miss every criterion. This spec targets failures observable only by measurement, along with the one rubric item that inspects implementation mechanics through browser DevTools (the bounded row check).

### C. *The tool configurations*

The following tool configurations were created for our custom agent. Each of these configurations would make different toolset available to the model, except for basic file tools (read, write, edit, list) which are available in every tool configuration.

1. *no_verification*: No tools other than the basic file tools are available to the model. The model must write everything correctly in one pass from its own knowledge. We refer to this configuration informally as building blind. The question it answers is how good is pure authoring, and what breaks without any feedback?
2. *static*: Adds a linter and a type checker (tools that read code and flag likely mistakes without running it). We refer to this configuration informally as the linter condition. The question it answers is whether feedback about code without executing anything helps at all?
3. *boot_check*: Adds the ability to install the declared dependencies, attempt to start the application's server, and report whether it began listening. This tool does no other execution. We refer to this configuration informally as boot check. The question it answers is how much of a value is just knowing whether the application starts?
4. *execution*: Adds a full shell and the agent can run any system command in its workspace, including installing packages, starting and stopping background processes, and running both frontend (user interface and the client code that runs on the user's browser) and backend (the server side that contains the main business logic). The question it answers is what does unrestricted runtime ability buy, and what does it cost?
5. *visual*: Adds everything in the *execution* configuration plus a screenshot tool that renders a URL in a headless browser (a browser runs invisibly, without a window) and returns the image to the model's vision input. The question it answers is does seeing the page add anything beyond running the code?
6. *visual_no_shell*: Adds screenshots and an app-launcher abilities but does not allow arbitrary command execution. Instead, the agent harness starts the app and takes the screenshot. In this tool configuration, the model can only look. The question it answers is whether the sight ability helps and if it is the seeing or the shell underneath it?
7. *delayed_verification*: This provides file tools only until the model first declares it is finished. At that moment a shell unlocks, and the model is asked to verify and fix before finishing again. The question it answers is how much of verification's value is iterative development versus a final debugging pass?

8. *behavioral*: This adds only a run-tests tool that executes the project's own test command. The model must write its own test suite. The question it answers is whether self-authored testing could be a substitute for granted verification?

### *D. Why a passive screenshot rather than browser automation*

Our screenshot tool is built on Playwright driving headless Chromium, the same library that can click, type, drag, and resize a live page. Our agent exposed only its rendering call to the model for this study. A model in the visual conditions (visual and visual_no_shell) passes a URL to the agent and receives an image in response, and nothing more. The passivity is therefore a deliberate restriction rather than a limitation of the tooling available to us.

We restricted it for three reasons. The first is isolation. Exposing the interaction API of playwright would have granted seeing and acting together, making the visual condition a combination of multiple abilities and making it difficult to determine which half earned any gain. That is the failure this study exists to avoid, and it is why visual_no_shell exists as well, to separate seeing ability from execution ability. The second is measurability. Every screenshot is a single call with a single argument, so we can count exactly what a model chose to look at, which is what makes the attribution analysis of Section IV-D possible. A browser-automation session has no such unit. Its actions could be logged and counted too, but every click both changes the page and shows the result, so the count would mix acting with looking. And if the model chooses to drive the browser from a script, a single call can hide fifty actions. Either way, we could no longer count how often the model chose to look, which is what Section IV-D needs. The third is prior evidence. Our earlier study granted this same library with its interaction capability intact and found it raised cost by 42 to 68 percent without improving reliability, so testing the narrower channel in this study was the more informative order.

The restriction has a real cost. Faults that only appear when someone clicks lie outside what this channel can reach, so our visual conditions understate what sight could achieve. The boundary was not merely theoretical. One model, gemini-3.1-pro, installed its own scriptable browser in three runs to click elements the granted tool would not, and in 21 further runs models stated the limit in their reasoning before working around it or falling back on reading the code.

### *E. Why the execution configuration does not trump the other configurations*

A natural concern about the tool configurations above is that the execution configuration provides full shell access, and a capable model could use that shell to rebuild almost all of the functionality in the configuration ladder such as running a linter, starting the application and probing it, or installing browser-automation tools to take screenshots. Our study design does not prevent this, and it does not need to, mainly for two reasons. First, granted versus self-built is itself part of what we measure. The boot probe, for example, is a guaranteed single tool call and a model that wants the same functionality without boot probe (boot_check) tool must instead find the entry point and manage the process itself, which costs tokens. So, if a model rebuilds a capability, its token count rises accordingly, and token usage is one of our measured outcomes. Second, self-built sight cannot reach the model's eyes. When the granted screenshot tool runs, the harness attaches the captured image to the model's next message as visual input, so the model sees actual pixels. A model that installs a browser and takes its own screenshot can only open the screenshot file with its file tools, which in our harness return bytes, not vision. Outside of the two sight conditions the model is called with no visual input at all, a setting each run's manifest records as vision = false. The nearest substitute available to a shell-holding agent would be to compute facts about the screenshot with a script, its dimensions or its pixel colors, which yields text about the page rather than sight and the command logs show no run attempted even that. Only the harness can place a picture in front of the model.

In practice, this concern turns out to be almost entirely moot. We searched every logged command of all 252 runs that had an unrestricted shell without granted sight (the execution and delayed_verification conditions). One model, gemini-3.1-pro, scripted a headless browser with Puppeteer in three of those runs, using it to drive drag-and-drop and reordering interactions on the kanban board but none of the scripts ever captured a screenshot, and no image was ever read back into the model's input. So, in practice, no model ever gave itself sight. The gift of sight remained ours to grant or withhold, and that grant is what the visual conditions measure.

### *F. Application and tool-configuration pairing*

The five core conditions (no_verification, static, boot_check, execution, and visual) were run on every application. The three targeted conditions (visual_no_shell, delayed_verification, and behavioral) were run only on specific applications (Table I), mainly to control the cost, as running every configuration on every application and model would have multiplied both the number of builds and the token cost.

TABLE I
APPLICATIONS AND TOOL CONFIGURATION PAIRINGS

| Application | core five | visual_no_shell | delayed_verification | behavioral |
|---|---|---|---|---|
| message-board | Yes | No | No | No |
| kanban-board | Yes | No | Yes | No |
| seat-booking | Yes | No | Yes | Yes |
| calendar-week-view | Yes | Yes | No | No |
| metrics-dashboard | Yes | Yes | No | No |
| kanban-labels-brownfield | Yes | No | No | No |
| log-explorer-perf | Yes | Yes | No | No |

The study contains 1,116 runs in total: 192 each in no_verification, static, boot_check, and execution. 174 runs in visual, 90 in visual_no_shell, 60 in delayed_verification, and 24 in behavioral.

Each combination of application, condition, and model ran four times by default, and a targeted subset ran a fifth time. The batch reached these counts in waves rather than one step. Every combination started with one run, and repeats were added while the batch was underway, guided by the token costs the harness records automatically and by an early trial scoring of roughly two dozen runs on two tasks (not included in the final run), used only to check that the tasks and probes were behaving and later superseded by the full grading of all 1,116 runs. The fifth run went where it could tell us something. Some of our questions ask how often builds fail, and with four runs a real but uncommon failure can easily appear zero times, so a fifth run gives it one more chance to show. Two conditions had token costs that swung widely between repeats, and one more run steadies their averages. The message-board was skipped since it was a simpler task and every model was expected to score near perfect on it under every condition, so an extra run there could reveal nothing new. Three safeguards keep this from tilting any comparison. Both sides of every planned comparison received the same number of runs, the analysis plan was frozen before the full condition-blind grading began, and rerunning the entire analysis without the fifth runs leaves all six primary verdicts unchanged, so no finding in this paper depends on them. The exact run count for every task and condition is listed in the released statistical output (Appendix F of stats-results.txt).

Also, not every application was paired with each tool configuration. visual_no_shell condition exists to separate sight from execution. It grants the screenshot tool but no shell, so any difference from the full visual condition is the value of running, and any difference from plain execution is the value of seeing. We ran it on the two visually hard tasks, the calendar-week-view and the metrics-dashboard, where a still image can catch the kind of failure that matters, such as an event drawn in the wrong place or an axis label that is clipped.

We also ran visual_no_shell on the log explorer, but this time as a negative control (a case arranged so that the effect we are looking for cannot be present). If the measurement still reports that effect, the measurement is what is at fault. The log explorer's failures are slow scrolling and a list that stutters under a hundred thousand rows, and a still screenshot cannot show motion or delay. A photo of a laggy list and a photo of a smooth one look the same. Sight therefore has nothing to catch here by construction, so if either screenshot arm were to beat plain execution on this task, the gain could not be real, and we would suspect our own instrument before believing it. The committed form of this check is the S3 contrast, visual versus execution on this task and the screenshot-only arm gives a second, descriptive reading of the same control. Both come out flat or slightly negative, the reassuring outcome, reported in Section IV-F.

delayed_verification exists to observe "finish first, then verify it works". This needs applications whose concurrency makes their runtime behavior complex enough to hide surprises, so it was run on the two hardest API tasks, kanban-board and seat-booking. The two visually hard tasks, calendar-week-view and metrics-dashboard, would not have been good candidates for since an unlocked shell still cannot see the visual failure modes those specs are designed to create, so the condition would not measure what it targets there.

The behavioral condition was run only on seat-booking. The behavioral condition gives the model only one checking tool, a command to run its own test suite. That tool produces no feedback unless the model first writes tests, so to benefit from it at all the model must author its own tests. That fits seat-booking well, so we ran the condition there. Seat-booking's hard parts are its race conditions and its hold-expiry logic, and both are exactly the kind of thing a developer would normally check by writing a test, such as firing many requests at one seat and asserting a single winner or letting a hold expire and asserting that the seat frees. We did not run the condition on the other tasks, for two reasons. On the calendar-week-view and metrics-dashboard, the failures are about how the page looks, and a test suite cannot judge rendered geometry, whether an event sits in the right place or an axis label is clipped. On the message-board, the easy floor, we chose not to spend the condition's runs where the ladder already covers the same ground.

### G. *Controlled test environment*

We wanted the environment to be as controlled as possible, so we took the following steps:

- *Identical text*: The base prompt template, the environment briefing, and each task spec are identical across conditions. We enforce this by fingerprinting. Each run's log records a cryptographic hash of every text component, and we confirmed that every run reported the same hash for the prompt template, the briefing, and the task text. The one thing that changes between runs is the workspace path. The base prompt is a template with a placeholder for the working directory, and each run fills in its own folder, so the final prompt the model reads is the same in every run except for that path. Each run records its workspace path, and the fully assembled prompt is hashed into the manifest as well. One could take the one published prompt template, fill in that run's recorded path, append the one published briefing, and hash the result and it would match the run's recorded hash. Because the only ingredient that changed was the workspace path, and every other piece came from the single published copy, a match proves nothing else was altered. We checked this reconstruction for all 1,116 runs.
- *Environment briefing*: The environment briefing tells every model, in every condition, that the listed tools are the complete set and that dependencies declared in the package manifests would be installed once the run finished. We added this after a few initial test runs (not part of the 1,116 runs for the final study) showed that some models ignored some tools unless told they were available. The briefing is identical across all runs and expresses no preference for any tool or configuration.
- *Sampling settings*: We fix temperature to 0.0 and top_p to 1.0 (the two settings that control randomness in a model's output, with temperature 0 requesting the most deterministic behavior available) for every model that accepts these parameters, with two caveats. First, some models, notably the Claude Opus family and gpt-5.5, reject explicit temperature and top_p. These models run with their provider's default sampling. This does not distort the comparison. When a model falls back to its provider's default sampling, it uses that same default in every one of its conditions, so the sampling setting never differs between the conditions we compare. It matters that we only ever compare a model against itself. We never read a result by placing one model next to another, only by placing a model's tools-on runs next to its own tools-off runs. Since the sampling setting is the same on both sides of that comparison, it cannot be mistaken for the effect of the tools. Second, temperature 0 does not guarantee identical outputs across repeats, because providers are not perfectly deterministic, which is why we run replicates (repeat runs of the same model, task, and condition).
- *Provenance and budgets*: The 1,116 runs that created the applications executed between June 12 and July 11, 2026, and every request from every model traveled through the same single API gateway, so no model was reached through different infrastructure than another. The exact model identifier sent with each request is recorded in that run's manifest, alongside its start and end times, step count, and token totals Every run ended with the model declaring completion and none was cut short by the harness or by any resource limit.

### H. *Analysis rules*

We encountered two situations in the study that required explicit handling rules.

*Situation 1: A model ignores a granted tool*

Granting a screenshot tool, or any other tool, does not guarantee that a model will use it. Some models took zero screenshots in the visual conditions. In these cases, we grade the application by the condition it was assigned, and not by the behavior the model exhibited. This is similar to the intention-to-treat rule from clinical medicine where a patient who never picks up a prescription still counts in the drug group, because silently dropping such cases would bias the comparison toward the runs that happened to behave.

Non-use was concentrated rather than spread. The three Claude models invoked every granted checking tool in all 462 of their tool-granted runs (154 each, across the seven conditions that grant any verification tool), while grok-4.3 alone accounts for 55 of the study's 100 non-use runs. It also depended on the tool. No model ever skipped the boot probe, every delayed_verification run used the shell once it unlocked, and only two runs in the study (both in the execution condition) finished without issuing a single command. But the linter went unused in 44 of the static condition's 192 runs (23 percent), the screenshot tool in 47 of the visual condition's 174 runs (27 percent) and in 3 of the 90 visual_no_shell runs, and 4 of the 24 behavioral runs never once ran the test suite despite the fact that it was their only checking tool. Models seemed to skip the checks they can finish without. Because every comparison is within-model, a gap measures the effect of granting a tool, not of using it. So a model that ignores its tool builds the same way with it as without it. The two sets of runs (with and without the tools) come out alike, and the gap between them shrinks no matter how useful the tool would have been. Section IV-D discusses this when attributing the sight effect.

*Situation 2: An application never starts*

If a model delivers an application that fails to boot, it is graded as a failure on all rubric items. Excluding non-starting builds from analysis would quietly reward the conditions that produce the most non-starting application through a selection effect. Suppose a weak condition only manages to start one build in three. Those few that do start are not a fair sample of its work as they are its luckiest attempts and the ones where the model happened to get enough right to boot. Drop the other two-thirds and we are averaging over this hand-picked best of its runs. On the other hand, a strong condition, where nearly everything starts, is instead averaged over almost all its builds, including the many that merely work without being excellent. Scoring only the survivors therefore compares a weak condition's best runs against a strong condition's typical ones, and the weak condition could come out ahead. Counting every non-starting build as a zero allows each condition to be judged on all the runs it was given. We also analyze startup failures as a separate class of outcome.

### *I. Tracking Tool usage*

Because we built the agent, every tool invocation is logged with its arguments, including calls the agent refused because the tool was not in that condition's list. From these logs we compute per-run tool-usage counts (internally referred to as tool_uptake), such as how many times a run called the linter, the boot probe, or the screenshot tool. These counts let us confirm that no run ever invoked a tool outside its condition's granted set, a check we ran over all 1,116 runs. Screenshot use in the visual condition, for instance, ranged widely, from runs that never called the tool to runs that called it more than twenty times.

### *J. Grading mechanism: Primary human grading, secondary API Probes*

Every application built during this study follows the same grading process. Because these are all web applications, the primary grading instrument is human grading. We chose a blinded human over an automated model judge deliberately. LLM-as-a-judge evaluation [21] scales well, but a judge model can share training lineage, and therefore correlated blind spots, with the very models under test, a human working against a frozen rubric keeps the measuring instrument independent of everything being measured.

Grading was condition-blind by procedure. Runs were graded in a mechanically shuffled order, fixed-seeded and never sorted by condition. Every scorecard carried no model or condition field and the grader did not consult a run's manifest or logs before scoring. Because the live run paths encode the condition, this blinding was procedural rather than enforced by concealment. The regrade described below also provides a direct check where the regrade pass was mechanically concealed and each sampled run staged under an opaque identifier with path, task, and condition redacted. The regrade process reproduced the first-pass scores at 98.8 percent item agreement with every survival call unchanged, so the procedural blinding measurably held.

Every application has a rubric card of weighted, human-scored criteria. The author worked through the 1,116 runs running each application and scoring each rubric item as pass, partial, or fail, with written notes wherever appropriate. Blinding was enforced by the setup, not left to discipline by grading runs in a shuffled order under opaque identifiers, with the model and condition fields stripped from every scorecard. This produced the human interface score, the one quality measure applied to every application in the study.

We also ran automatic probes as a secondary check on the subset of applications whose target behavior was observable through the API. Each probe run staged a fresh copy of the application in a separate area. Copies that failed to boot were recorded as boot failures and kept their zero scores in every average score calculation, for the same reason as Situation 2 in Section II-H i.e. setting them aside would hide exactly the failures the study is designed to count. The rest were subjected to weighted API probes that sent API requests to the backend to stress the application in a way hostile or unlucky real usage would. For example, one such API probe posts a message and watches two connected clients both receive it. Another kills the process without warning, restarts it, and checks that the data survived. A third fires eight simultaneous hold requests for one theater seat and requires exactly one winner. The result is a functional score, the percentage of automatic probe weight passed.

Three applications have no API probes at all, and this is by design. The calendar-week-view's and metrics-dashboard's difficulty is layout, and the log explorer's is measured smoothness, none of which an API probe can observe. For these, the human rubric score is the only grade.

The two grading mechanisms, human grading and automatic probing, also cross-check each other. Every run the machine recorded as a boot failure was examined by hand as well. The human grader either confirmed the failure or got the application running through a launch path its own files declare. A confirmed failure would then result in “fail” score on every rubric item. A run that the grader could start keeps its human grade and counts as surviving.

For survival we go one step further and we compute it twice, by two routes that share no data. The first route uses the machine's record. Did the automated grader boot the backend, corrected by the hand audits (the ten builds whose server ran but whose page never loaded, and the few the machine could not start but a human could through a setting the app itself declares). The second route ignores the machine entirely and reads only the human's scorecard. An app that earned any points must have been running in front of the grader, so it survived; an all-fail card counts as surviving only when the audit note says the app came up and failed on its merits rather than never coming up. Two independent routes targeting one question agreed on all 1,116 runs. A released check, *verify_post_merge_results.py*, recomputes both routes and reports any disagreement, so the survival numbers rest on neither the machine nor the human alone.

All 1,116 human grades come from a single grader, so we measured the stability of that instrument directly with a blind regrade. Four weeks after grading closed (grading ended July 15 and re-grading ran from August 16 to August 19, 2026) and we regraded a 10 percent sample i.e. 112 runs drawn with a fixed, released seed, stratified proportionally across task and condition. 18 runs whose records had been re-opened during manuscript writing were excluded from the sampling frame, and runs that never launched remained in the re-grading sample. The samples were re-staged under fresh opaque identifiers in a new shuffled order, with model and condition stripped and the original scorecards out of view, and scored against the same rubric and accommodation rules. The agreement between the two passes is as follows. The per-item exact agreement 98.8 percent (589 of 596 weighted rubric-item comparisons), linearly weighted kappa 0.973 [22], [23], intraclass correlation ICC(A,1) [24], [25] 0.997 on the per-run score, and a mean absolute per-run difference of 0.52 points on the 100-point scale which is "almost perfect" by the Landis-Koch convention [26], despite 93 percent of item verdicts on launched runs being "pass," a prevalence that makes chance agreement high [27]. Because a non-launching run scores all-fail by rule, we report the launched-only split alongside the pooled figures, which are as follows. 98.7 percent exact agreement, weighted kappa 0.940, ICC 0.991. Every item-level disagreement between original and regrade scoring pass was between adjacent categories (pass/partial or fail/partial) and none crossed the full pass/fail distance. Most importantly, the two passes agreed on the survival call for all 112 runs, and no run moved between a zero and a non-zero score, so the survival measure of Section II-L is stable under re-measurement. However, the regrading aims to verify repeatability, not correctness. The full protocol for regrading was written before the sample for re-grading was drawn, the seeds, the sealed sample and mapping, every second-pass card, and the agreement analysis are released in the archive under the regrade folder.

### *K. Launch accommodations and the configuration-space rule*

Some artifacts declared development launch scripts that cannot launch their own user interface even though the application itself is sound, such as a script that starts the frontend toolchain in a way that skips the artifact's own configuration file, a proxy rule that intercepts requests for the application's own code, or a prebuilt bundle compiled against a port the server does not use by default.

When the human grader encountered these, the governing rule was that an accommodation may only select values or paths within the artifact's own declared configuration space (scripts it defines, configuration files it ships, and environment variables its own code reads, such as setting PORT for a server that explicitly honors it) and may never add information the artifact does not contain. If some declared combination renders the interface, it is then graded on its own merits and the configuration defect is recorded in the run's notes.

If no declared combination works, the render-dependent criteria fail. These configuration inconsistencies are genuine model-authored defects. But the rubric has no criterion for launch mechanics, and we did not fold them in the other existing rubric items for these mainly for three reasons. First, the scorecard was written and frozen before a single run existed, and none of its items asks how the application launches. Adding a launch penalty after seeing the results would mean changing the yardstick mid-study. Second, we did not find these defects evenly. They surfaced only in the runs where launch trouble forced the grader to dig, and they concentrate in particular models' habits, so scoring them would punish the runs we happened to catch, spare the ones we did not, and land hardest on some models for no better reason than where we looked. Third, a fair score needs one measurement applied to every run alike, and no instrument or rubric in the study checked for "the app's own declared start command works from start to finish" identically for all 1,116 builds. The defects are not lost. Section VI-A measures the two most common classes of launch defects mechanically, over every run, where they can be counted fairly.

### *L. Whole Application survival*

Survival in this paper means the application as a whole comes up through at least one path that its own files declare. The automatic boot check observes only the backend. Hand grading revealed a complementary failure class, builds whose backend starts cleanly while the frontend can never render, because the frontend source itself is broken (in every observed case there was a frontend file with a parse error, so no declared path, dev or production, could build or serve the page).

Throughout the study we follow one pattern, whenever hand-grading surfaces a recurring problem. We write the observation down, turn it into a mechanical check, and re-apply that check identically to all 1,116 runs, so a discovery made on some runs never becomes a judgment applied to only those runs. Following that pattern, the human grader's observations became a batch-wide audit. We pulled every run where the two instruments disagreed. For example, the automatic probe had booted a run's backend successfully, yet the human's scorecard had come out all fails. Each of these runs was re-examined against the notes the human had written while grading blind. Ten of them turned out to be genuine frontend launch failures, builds whose server ran but whose page could not be brought up through any path the build itself declares. Those ten are recorded in a released audit file (frontend-launch-audit.csv) and counted as non-survivors. Applications that render but still fail their rubric criteria remain survivors. Not launching and launching badly are different failures. By contrast, an application that never starts scores fail on every rubric item rather than being skipped, again to avoid survivorship bias.

Another automatic score, architectural fidelity, checks by static scan that each build used the mandated stack rather than substituting another database or transport. The scan awards eight points. Two of them are for the realtime channel, the stack's live-update mechanism (server-sent events) with one point for a server that sends the event stream, and one for a client that listens to it. Four of the seven applications require live updates, and every one of their builds scored a perfect 1.0. The other three (the calendar-week-view and the metrics-dashboard, and the log-explorer-perf) have no live-update requirement, so those two points cannot be earned there, and their maximum is six of eight points, 0.75, and every one of their builds reached it. No build anywhere swapped in a forbidden substitute (such as sqlite3 for the database or websockets for the transport), so the verification tools did not change which technology stack models built with.

## III. HYPOTHESES

We finalized and documented six primary hypotheses in a written analysis plan (stats-plan.md, released in the archive), each a specific prediction about how changing the verification surface should change an outcome. The plan was committed, in writing and before any statistic was computed, which comparisons count, how each is tested, and what qualifies as support. One caveat that the plan itself records is that it was finalized after grading was complete, so descriptive averages had been seen and what was fixed before any inferential computation is every remaining choice i.e. the tests, the correction, the handling rules, and the support criteria. We therefore describe the hypotheses as pre-specified rather than pre-registered in the external-registry sense. Each hypothesis is tested with one primary comparison between conditions.

1. *P1 - more verification helps*: An agent that can run and exercise its own code produces better working software than one with no checking tools at all. We compare the execution condition against the no_verification condition on the functional (behavioral) score, on the three API-checkable applications (message-board, kanban-board, and seat-booking). The execution condition grants a shell and no sight, so in practice it reaches the application through the API, the server log, and the process rather than the rendered page. The functional probes measure that same surface, and they measure it more severely than a human grader could, such as firing eight simultaneous hold requests at one seat and requiring exactly one winner, killing a process without warning and checking that data survived, and confirming that two connected clients both receive a message.
2. *P2 - the cheapest check*: Even the smallest verification in the ladder should change what ships. The boot probe reports a single bit, did the server start, and repairs nothing itself. The hypothesis is that this bit is enough that models that build blind and ship launch failures, could have fixed the launch failures had they known about them. So, a bare report of a failed start should let them catch and repair those failures before finishing, making far more builds launchable. We compare the boot_check condition against the no_verification condition on survival, across all seven applications.
3. *P3 - verification has a cost*: Running and iterating is not free and a full shell should consume more tokens than working blind. We compare the execution condition against the no_verification condition on token cost, across all seven applications.
4. *P4 - does sight help*: On the tasks whose hard part is visual, letting the agent see a screenshot of the page it built should add quality beyond what a shell alone provides. On the two visually hard tasks (the calendar-week-view and the metrics-dashboard), we compare the visual condition against the execution condition on the human interface score.
5. *P5 - modifying versus writing*: It is natural to expect verification tools to matter even more when the agent changes code it did not write, because unfamiliar code is easier to break. Our analysis plan did not commit to that direction. It tested for a difference either way, and recorded our expectation that the difference would be near zero. On this hypothesis, support means a confidence interval close to zero, not merely a non-significant permutation p-value. We test whether the tool effect differs between the modification task (kanban-labels-brownfield) and the equivalent new-build task.
6. *P6 - measurement helps where performance is the challenge*: On a task whose failures are about measured performance, such as user interface staying responsive over a very large dataset, giving the agent a shell that can run and measure the application should help more than giving it no tools. On the performance task (the log-explorer-perf, which must stay smooth over 100,000 rows), we compare the execution condition against the no_verification condition on the human interface score.

TABLE II
PRIMARY HYPOTHESES: PERMUTATION TEST RESULTS (HOLM-CORRECTED ACROSS THE FAMILY OF SIX)

| Hypothesis | Contrast | Outcome | Effect | 95% CI | perm p | Holm p | Supported |
|---|---|---|---|---|---|---|---|
| P1 | execution − no_verification | Functional | +12.26 | [+7.04, +17.47] | 0.0001 | 0.0006 | Yes |
| P2 | boot_check − no_verification | Survival (0-1) | +0.13 | [+0.10, +0.16] | 0.0001 | 0.0006 | Yes |
| P3 | execution − no_verification | Log tokens | +0.80 | [+0.73, +0.87] | 0.0001 | 0.0006 | Yes |
| P4 | visual − execution | Human | +6.88 | [+0.83, +13.41] | 0.0413 | 0.0826 | No |
| P5 | (brownfield gap) − (greenfield gap) | Functional | -0.67 | [−13.67, +12.67] | 0.9989 | 0.9989 | No |
| P6 | execution − no_verification | Human | +15.33 | [+7.33, +24.00] | 0.0225 | 0.0675 | No |

TABLE III
SECONDARY CONTRASTS: PERMUTATION TEST RESULTS (UNCORRECTED, NO HOLM)

| Hypothesis | Contrast | Outcome | Effect | 95% CI | perm p |
|---|---|---|---|---|---|
| S1 | execution − boot_check | Functional | +2.41 | [−0.38, +5.21] | 0.1427 |
| S2 | visual_no_shell − execution | Human | +4.98 | [−1.42, +11.75] | 0.1951 |
| S3 | visual − execution (log explorer) | Human | −2.17 | [−12.50, +7.33] | 0.8571 |
| S4a | static − no_verification | Functional | −3.60 | [−11.36, +3.61] | 0.3886 |
| S4b | static − no_verification | Survival (0-1) | −0.02 | [−0.07, +0.03] | 0.3534 |
| S5 | behavioral − execution | Functional | −16.07 | [−31.67, −1.70] | 0.0425 |
| S6 | visual − execution, per model | Human | Per-model | | |

The six hypotheses above (P1 to P6) are the primary contrasts. Each was decided in advance in our analysis, and together they form one family. We perform permutation tests on these (a permutation test asks how often a difference as large as the observed one would arise if the condition labels were shuffled at random) [28], [29], the standard way to compare systems in language-processing research [30], [31], calculate permutation p-values, and then apply the Holm correction [32] across all six (testing six hypotheses gives chance six opportunities to produce a false positive, so Holm raises the bar accordingly). A primary result is called supported only if it survives that correction. These carry the paper's main claims. Their results appear in Table II.

The secondary contrasts are additional comparisons that decompose or probe the primary results rather than test a new headline claim. Because they are exploratory rather than pre-specified headline tests, we report them without the Holm correction and read them as suggestive. They can help explain why a primary result looks the way it does, but on their own we do not treat them as confirmatory.

The secondary contrasts we report on are(results in Table III):

- *S1* - whether a full shell adds anything beyond the cheap boot probe (execution versus boot_check). The boot probe is the cheapest verification in the study, and the shell costs nearly three times as much, so this contrast prices what the shell's extra functionality actually buys. A large gap would justify the shell's premium. A small one would make the probe the better bargain for most work.
- *S2* - whether sight helps once it is separated from the shell (visual_no_shell versus execution). The visual condition mixes two channels, running and seeing, so its gains could come from either. The visual_no_shell condition keeps the screenshots and removes the shell (Section II-F). If visual_no_shell still keeps up with execution on the visually hard tasks, then whatever sight contributes, comes from the seeing itself and not from the shell underneath.
- *S3* - whether sight adds anything on the log explorer, whose failures can be measured but not seen (visual versus execution). Because a still screenshot cannot show a lag or stutter, we predict no benefit, which makes this contrast a negative control (Section II-F). A flat or negative result is the reassuring outcome. A benefit would point to a problem in our setup rather than a real effect.
- *S4* - the linter's null result (no measurable effect) in more detail (static versus no_verification, on both functional score and survival). Linters are cheap and widely assumed to help, so a null is worth pinning down rather than asserting. This contrast measures the linter's effect twice, once on behavioral quality and once on whether builds start at all. No improvement on either would mean that granting a tool is not the same as benefiting from it.
- *S5* - whether self-authored tests substitute for a granted shell (behavioral versus execution). The behavioral condition's only checking tool runs tests which the model writes for itself (Section II-F). This sounds like it could deliver the shell's benefits from a much smaller surface. Parity here would mean a small, targeted tool can replace the full shell. A shortfall would suggest that, under a fixed budget, building a test harness competes with finishing the application itself.
- *S6* - a per-model view of the sight effect (visual versus execution, split by model). A model that is given a tool may never use it. And an average pooled over all the models cannot say where an improvement came from, a gain from actually looking at screenshots and a gain from anything else would look identical inside the average. This split places each model's gain next to how often it actually took screenshots. If the gains concentrate in the models that looked most, sight works the way it is supposed to. If they land on models that barely looked, the pooled benefit cannot be credited to looking.

## IV. RESULTS

This section reports what changed when we changed the verification surface, and every reported number is built in two steps.

In the first step, we work within each model on each task. For every model-and-task combination (say, gpt-5.5 building the seat-booking app), we calculate difference of averages by comparing that model's runs under one condition against its own runs under another and take the difference (for example, its average execution score minus its own average

no_verification score on that task). Differencing a model against itself in this way normalizes away that model's own baseline behavior, so a strong or weak model cannot distort the comparison.

In the second step, we combine these differences across all the model-and-task combinations into the single figure we report. That combination is a weighted average. Each combination counts in proportion to how many runs it contains, so a combination with more runs contributes more and none is over or under counted. So, a statement like "the execution condition scored 12 points above the no_verification condition" is a within-model, within-task difference first, and then a run-weighted average across every model-and-task combination. The reported figures therefore do combine all models. They simply do so after each model has been measured against itself, which is what keeps the comparison fair. Where the per-model picture differs from this pooled view, we show it explicitly.

Before computing any statistic, we wrote an analysis plan that fixed which hypotheses we would test, how we would test them, and the rules for handling awkward runs (Section III). The tests themselves are stratified permutation tests. They ask how often a difference this large would appear if the condition labels were shuffled at random, with shuffling allowed only among runs of the same model and task, so that model or task differences are not mistaken for tool differences. Because we test several hypotheses, we apply the Holm correction to the p-values to guard against false positives. We also computed a 95% confidence interval for each hypothesis by stratified bootstrap [33], to show how large each effect is and how much uncertainty surrounds it.

Each permutation test uses 10,000 permutations [28], [29], [30], [31]. The 95 percent confidence interval is calculated by rebuilding the dataset 5,000 times with redrawing runs at random from within each model-task-condition cell, and watching how much the effect varies across those rebuilt datasets (a stratified bootstrap, the interval covers the middle 95 percent of that variation). The random seed (20260703) is written into the released analysis code, so rerunning it reproduces every number in this paper exactly. Alongside each contrast, the released output also reports a simpler cross-check that assumes nothing about how the scores are distributed: a sign test, which asks whether the six per-model differences lean in one direction more consistently than six-coin flips plausibly would.

As a cross-check, we analyzed the same six contrasts again using a completely different statistical method, a mixed-effects regression [34] (Table IV). This method fits one model per contrast. It gives each of the six models its own baseline level, so that runs from the same model are not treated as unrelated to one another, and it adds a term for each application whenever a contrast covers more than one. The two methods agree. Every primary-effect points in the same direction and is the same size, matching the permutation estimates to two decimal places. The regression p-values, which are two-sided and uncorrected, lead to the same qualitative conclusions, including that P4 and P6 remain suggestive rather than established. The complete output is released as *stats-mixed-effects.txt* under *paper/* in the repository cited in the Data and Code Availability section.

TABLE IV
MIXED-EFFECTS CROSS-CHECK OF THE SIX PRIMARY CONTRASTS

| Hypothesis | Contrast | Outcome | Estimate | SE | 95% CI | p (two-sided) |
|---|---|---|---|---|---|---|
| P1 | execution − no_verification | Functional | +12.26 | 3.41 | [+5.57, +18.95] | .0003 |
| P2 | boot_check − no_verification | Survival (0–1) | +0.13 | 0.02 | [+0.09, +0.17] | <.0001 |
| P3 | execution − no_verification | Log tokens | +0.80 | 0.05 | [+0.70, +0.91] | <.0001 |
| P4 | visual − execution | Human | +6.88 | 3.80 | [−0.57, +14.32] | .0705 |
| P5 | (brownfield gap) − (greenfield gap) | Functional | −0.67 | 7.94 | [−16.24, +14.90] | .9331 |
| P6 | execution − no_verification | Human | +15.33 | 6.99 | [+1.63, +29.04] | .0283 |

The full plan, the analysis code, and the complete statistical output, including per-cell means, standard deviations, and counts, are in the released archive (*stats-plan.md*, *stats_analysis.py*, and *stats-results.txt* under *paper/* in the repository cited in the Data and Code Availability section).

Having run the pre-specified tests, we report the findings below. Each subsection focuses on one question, whether about a particular tool, a particular kind of task, or the cost of verification. Table V maps each subsection to the hypotheses behind it, listing the primary contrast and any secondary ones, to help trace every claim we make back to the specific test that supports it.

TABLE V
WHICH RESULTS SUBSECTION SUPPORTED BY WHICH HYPOTHESIS

| Subsection | Primary | Secondary |
|---|---|---|
| IV-A More verification, better outcomes | P1, P2 | S4a, S4b |
| IV-B The boot probe | develops P2 | S1 |
| IV-C Cost and variance | P3 | |
| IV-D The sight boundary | P4 | S2, S3 |
| IV-E Brownfield modification | P5 | |
| IV-F Performance budgets | P6 | |
| IV-G Self-authored tests | | S5 |

### *A. More verification, better outcomes, mostly at the bottom*

We measured three outcomes (survival, machine-probed behavior, and the human-judged interface) across the five core conditions (which ran across all seven applications). Every outcome improves as verification increases, and the biggest single jump comes at the very bottom of the ladder, the jump from no way to run anything (no_verification) to a single boot probe (boot_check).

*Survival*: Survival means the application as a whole comes up through at least one path its own files declare. Survival numbers for the five core conditions, also shown in Fig. 1, are: no_verification 86.5 percent (166 of 192), static 84.4 percent (162 of 192), boot_check 99.5 percent (191 of 192), execution 99.0 percent (190 of 192), and visual 99.4 percent (173 of 174). The failures that remain are real build failures, not grading accidents. We re-examined and classified each of the 66 failures (pooled across all conditions), and 56 of them come from the two conditions that cannot run anything (no_verification and static). Most of these failures trace to two causes.

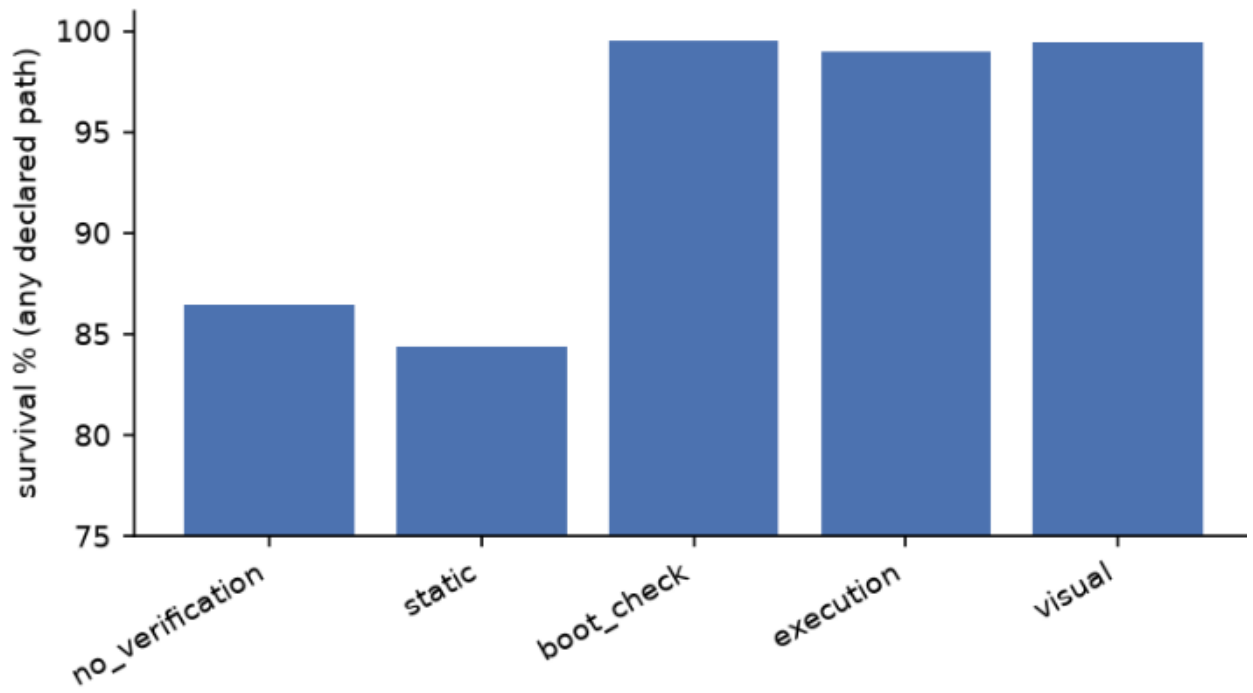

FIGURE 1. Launch survival across the five core tool configurations (all runs, intention-to-treat). Survival means the application as a whole, frontend included, comes up through at least one path its own files declare. The two conditions that cannot execute anything, no_verification and static, lose roughly one build in six to seven. A single boot probe removes all but one failure in 192.

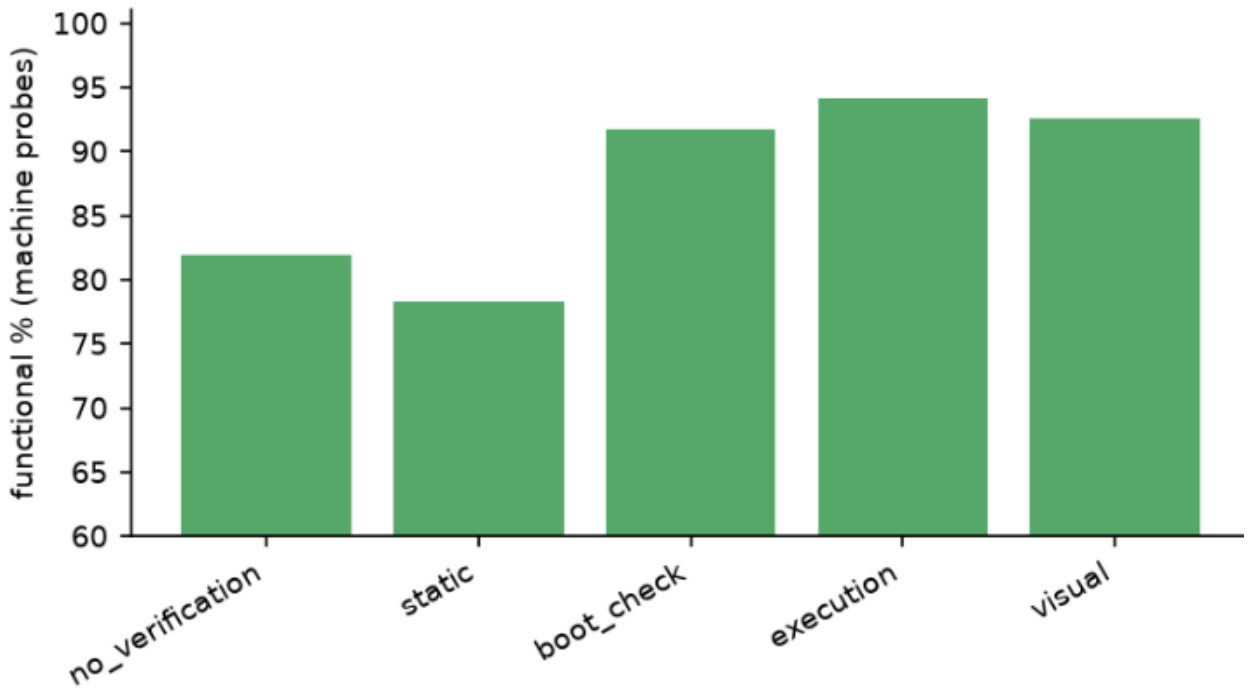

FIGURE 2. Mean functional score on the API-probed applications, by condition (all runs, non-starters scored zero). The largest step is from the no-execution arms to the boot probe, the shell arms add depth on the race, restart, and ordering probes.

First, there were source files that the model corrupted while editing, a botched splice that leaves the file unparseable, which any attempt to run or render the code would have caught at once (there were 35 backend cases and all 10 of the audited frontend cases). Second, there were database seed scripts that crashed the application on first boot.

The static condition, the one equipped with a linter and a type checker, did no better than having no tools at all. Its builds started slightly less often and scored slightly lower on behavior, and both differences are within statistical noise. The linter made no measurable difference, and that absence is itself a finding, because linters are cheap, familiar, and widely assumed to help.

The frontend audit adds a harder fact. The builds that had the linter were not even protected from the one mistake a linter is built to catch. Ten builds in the study shipped a frontend that could never render, in every case because of a parse error in a frontend file. Five of those ten were encountered in the static condition. Catching parse errors is precisely the job a linter exists to do, so the one condition equipped to prevent this failure produced half of its cases. Section V shows how that happens in practice, in a build whose model called the linter once, got an error back because it had never wired the tool up, and finished without ever checking again.

*Functional scores on the API-probed applications*: These are the machine-verified secondary checks. They pool the three fresh-build applications whose target behavior can be tested over the API, the message-board, the kanban-board, and the seat-booking. The modification task, kanban-labels-brownfield, also has probes, but those are reported with its own comparison in Section IV-E. Mean functional score across the five core conditions are: no_verification 81.9, static 78.3, boot_check 91.7, execution 94.2, and visual 92.6 (Fig. 2).

Our main planned comparisons (Table II)

1. execution versus no_verification: +12.3 points, 95 percent confidence interval [+7.0, +17.5], corrected $p < .001$, and positive in five of the six models. This means execution ability bought real quality and a model that could run its own code shipped software that behaved about twelve points better under our probes than the same model building blind. This is a gain too large and too consistent to be chance.
2. boot_check versus no_verification: survival effect is +13 percentage points, 95 percent confidence interval [+10, +16], corrected $p < .001$, concentrated in the models that fail without the boot_check probe (gemini gains 50 points of survival, gpt-5.5 gains 20, and the rest were already at ceiling, that is, near 100 percent with no room left to improve). This means the boot_check probe's single bit of feedback i.e. does the server start, was enough to rescue nearly all would-be launch failures, and the benefit went exactly where it was needed, the models that already launched reliably had nothing to gain, while the failure-prone ones gained the most.

A secondary contrast that we wanted to explore was *static* vs *no_verification*. The static condition effects are indistinguishable from no_verification on both outcomes. The functional observed score difference between the two conditions was −3.6 percentage points, 95 percent confidence interval [−11.4, +3.6]. On survival criteria the difference between the two conditions was −2 percentage points, 95 percent confidence interval [−7, +3].

*Human interface scores.* Every application in the study received a human interface score (Section II-J), and the released tables carry the ladder for all seven tasks. However, as per the pre-specified hypothesis P4, we focus on the two tasks whose difficulty is visual and for which the rubric is the only grade i.e. the calendar-week-view and the metrics-dashboard (the third rubric-only task, the perf-log-explorer, has its own section IV-F). On these two tasks, the human scores repeat the pattern of the survival and functional ladders i.e. quality climbs as verification grows, with the largest step at the bottom but the gaps between conditions are wider: this ladder runs from 72.6 (static) to 96.1 (visual), a spread of about 23 points, where the functional ladder spanned about 16. The complete human interface scores for these two tasks are: no_verification 72.8, static 72.6, boot_check 89.5, execution 89.2, and visual 96.1 (Fig. 3). In these numbers, two non-execution conditions (no_verification and static) sit 17 points below everything else, the boot_check recovers most of that gap, and the visual condition sits on top. Section IV-D examines how much of that top step can be credited to sight. These values are plain averages. For each condition we take the mean human score across all runs, pooled over all six models, so they show roughly where each condition landed. They are not the within-model comparisons we use to test the hypotheses, which instead measure each model against itself.

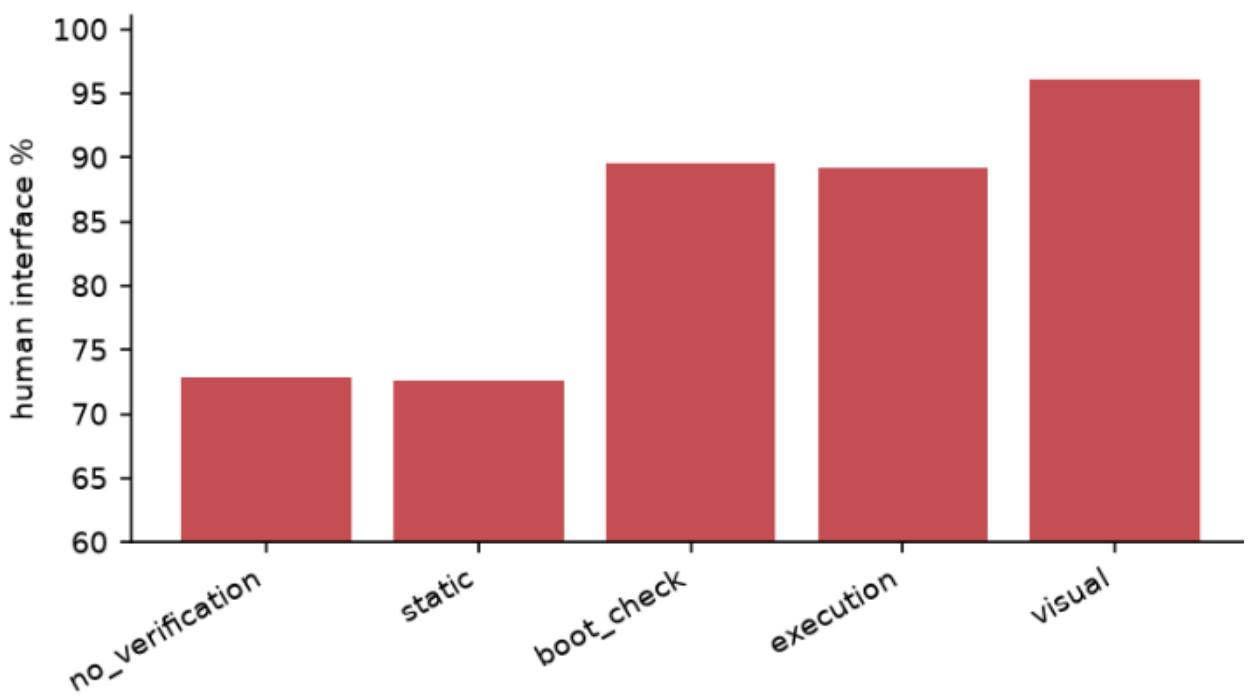

FIGURE 3. Mean human-graded interface score on the two visually graded tasks (calendar and dashboard), by condition. The same ladder as Figures 1 and 2 with a wider spread. Section IV-D examines how much of the top step is attributable to sight.

Reading all the three ladders together (survival, functional scores on the API-probed applications, and human interface scores) together shows what each step of verification buys. Survival improves first and improves the most. The boot probe alone lifts survival rates from 86 to 99.5 percent, a gain of 13 percentage points (Fig. 1 and the P2 row of Table II), and nothing after the boot probe improves it further. Behavioral quality comes next, and most of its gain also arrives with the boot probe.

Of the twelve functional points that separate blind builds from shell builds, ten arrive with the boot probe alone (Fig. 2). The shell adds the last two points or so, and they appear only on the most demanding probes. Those are the ones that fire several requests at the same instant to compete for the same resource, that kill the application without warning and restart it to check the data survived, and that verify items stay in the right order while the app is under load. Not every probe counts equally in the functional score. The automatic grader weights each probe, and eleven probe types carry the top weights of 3 and 4, the simultaneous-request races, the kill-and-restart persistence checks, and the ordering-under-load tests. Restricting to those probes across all four probe-bearing applications, counting every arm that ever held a shell (execution, visual, and delayed_verification) and only builds that started, shell condition passed 92.0 percent of that weight, the boot probe 91.1 percent, and the blind conditions (i.e. no tools or a linter only) 89.1 percent.

The human-judged quality of the page is the only measure that keeps improving past the shell. On the two visually hard tasks it climbs from 72.8 with no tools to 89.5 with the boot probe, 89.2 with a shell, and 96.1 with screenshots (Fig. 3). That final gain, screenshots measured against the shell, is the one result that does not survive our correction for testing six hypotheses at once, and Section IV-D sets out why.

Taken together, the three gradings tell one story with three different endings. Every measure improves as verification grows, but each stops improving at a different rung of the ladder. Survival stops first. The boot probe lifts the survival rates from 86 to 99.5 percent and nothing above the boot probe moves it again, because once nearly every application starts, a stronger tool has no launch failures left to prevent. Machine-probed behavior stops next. Ten of the twelve points that separate blind builds from shell builds also arrive with the probe alone, and the shell contributes only the small remainder, earned on the demanding probes that stress concurrency, restarts, and ordering. Human-judged quality is the only measure still climbing above the shell, and that final gain, screenshots measured against a shell, is also the only one in this section that does not survive our correction. The linter improves in none of the three. Each measure improves only while the added tools can still reach the defects it counts.

One number goes against the ladder. On the calendar-week-view, the boot_check condition's human score (96.6) beats the full shell's (86.4), even though the shell can do everything the boot_check probe can. The boot_check probe did not help more but the shell's runs on this task simply swung far more widely (a standard deviation of 31.0 against the probe's 3.3). Two models pulled the shell average down. grok-4.3 (48.2) and claude-4.6-sonnet (74.1) shipped their worst interaction defects in their shell builds on this task, while their boot-probe builds happened not to. Section IV-D returns to these two models.

### B. *The boot probe: the cheapest useful verification*

We measured how much the study's cheapest verification tool helps, and what it cannot catch. The cheapest tool is the boot probe. Its builds used the fewest tokens of any condition in the study, fewer even than builds made with no tools at all (Section IV-C lists every condition's cost). The short answer is that the probe delivers most of verification's value at that lowest price. It nearly eliminated launch failures and came within a few functional points of the full shell on the easy and medium tasks, at roughly a third of the shell's token cost. Though, the probe only knows whether the server starts listening. It cannot see what the server answers once real requests arrive, and it cannot see the page. The section below shows the limits of boot probe. On the seat-booking's demanding concurrency probes, on the metrics-dashboard's layout, and on the log explorer's measured smoothness, the probe's builds trail the shell's every time.

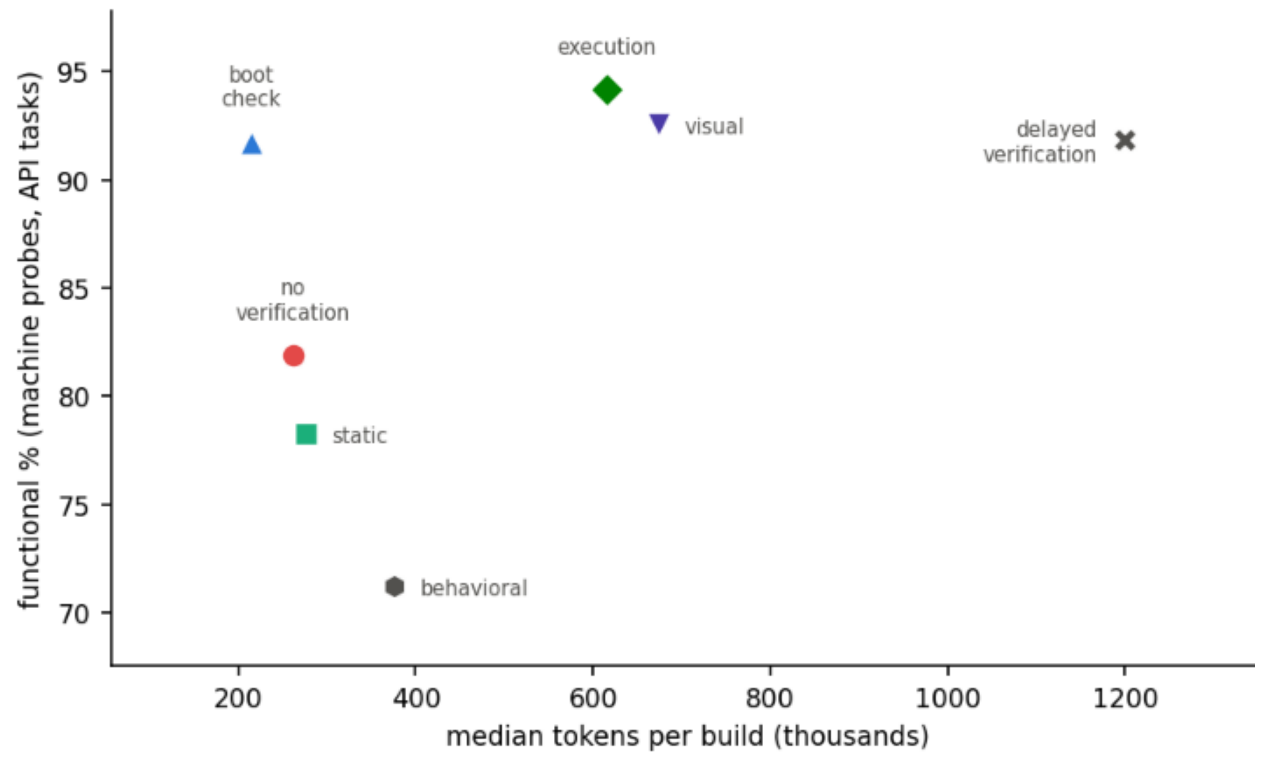

FIGURE 4. Median token cost against the machine-probed behavioral score (API tasks, all conditions). Each point pools the six models, which is consistent with the within-model design because every model contributes to every condition equally. The boot probe sits on the efficient frontier: cheaper than building blind and within a few points of the shell.

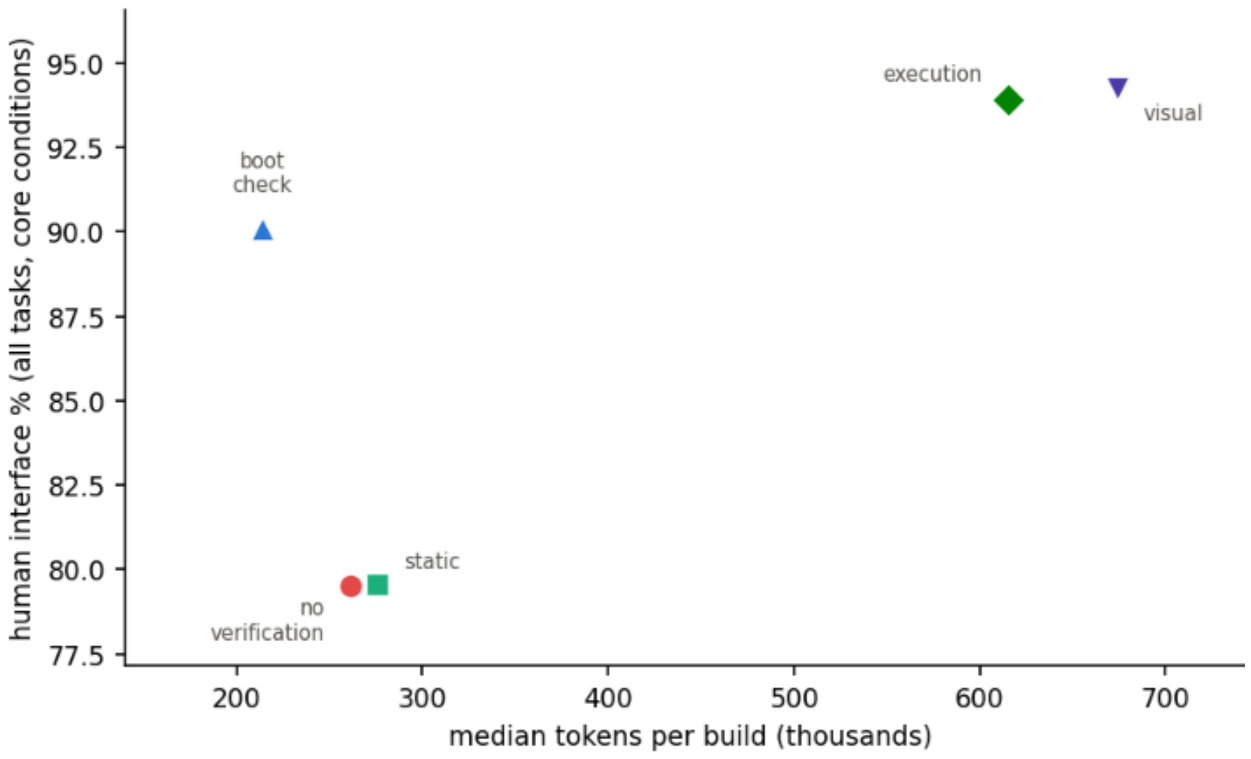

FIGURE 5. Median token cost against the human-graded interface score (all seven tasks, core conditions only, since the targeted arms ran on task subsets and would not be comparable). Machine and human judgments are plotted separately by design and never combined. The frontier's shape is the same under either judge.

The boot_check condition has one launch failure in the study (191 of 192 runs came up) at 214 thousand median tokens per build (Fig. 4), against the execution condition's 615 thousand. Its functional score (91.7) sits within 2.5 points of the execution condition's (94.2) on the on the three API-checkable tasks (message-board, kanban-board, and seat-booking). Yet its boundary is crisp. "Listens" is not "works." The boot check probe confirms that the server came up and accepts connections. It says nothing about whether anything the application does after that is correct.

The one build the boot_probe failed to save shows the blind spot exactly. Its backend started and listened, so the probe was satisfied, but a parse error in a frontend file meant the page could never be rendered.

A probe that only watches the server side cannot see such a failure. Two graded examples (Section V) show the same boundary at request time. An SQL type error that crashes the summary API endpoint on first call, and a permanently false error banner, both passing straight through boot probe because it only asks whether the server started.

On the two visually hard tasks (calendar-week-view and metrics-dashboard), boot_check's human interface scores trail the execution (Table VII) condition where layout is hard (82.5 versus 92.0 on the dashboard), and on the measurement task it trails by 9 points (84.8 versus 93.8). The boot probe buys survival, not correctness. The secondary contrast (S1 in Table III) quantifies this boundary. On the three API-checkable application, boot_check's functional shortfall against the execution condition is 2.4 points in the execution condition's favor, 95 percent interval [-0.4, +5.2]. So, on those tasks the shell's advantage in boot_check probe measured behavior may be nothing at all and is at most about five points, and it comes at 2.9 times the token cost. Where the difficulty is visual or has to be measured, the gap is larger, around nine points of human score on the dashboard and the log explorer.

## C. *Cost and variance*

We measured the token cost of each condition and how predictable that cost is from run to run. We found that the cost rises steeply as the verification surface grows, and the boot probe is the only configuration cheaper than building blind.

Median tokens per condition (all 1,116 runs) are: no_verification 262k, static 276k, boot_check 214k, execution 615k, visual 674k, visual_no_shell 659k, delayed_verification 1,199k, and behavioral 375k. Granting a full shell (execution) multiplies the no_verification baseline's cost by 2.35. Adding screenshots (visual) costs about a tenth more than the shell (execution). visual_no_shell costs about the same as visual, so sight without a shell is no cheaper. The declare-done-then-discover workflow (delayed_verification) is the most expensive configuration in the study, at nearly double the shell (execution). And boot_check is the only configuration that costs less than building blind (no_verification).

That last point is worth explaining, since adding a tool would normally add cost. The saving comes not from writing less code but from re-reading less conversation. The conversation is re-sent to the model on every step, so input tokens dominate cost, and an agent that takes fewer steps pays less to re-read its own session. Boot-probe runs finish in fewer steps than blind runs (a median of 17 against 19) and re-read far less (195 thousand input tokens against 247 thousand) while writing only somewhat less code (13.5 thousand output tokens against 16.8 thousand). Nor is the saving an artifact of cheap failures. Blind runs that never started cost less than blind runs that worked (182 thousand tokens against 263 thousand), yet comparing only builds that started, the boot probe still costs about 49 thousand tokens less than working blind. This is descriptive rather than tested, and it is consistent with the boot probe also having the lowest run-to-run spread in the study.

Now looking at the most expensive tool configuration (delayed_verification) most of its extra cost goes on a repair cycle. Once the shell unlocked, every one of the 60 runs used it, and 45 of the 60 went on to change their own code before declaring completion a second time. So, most of that extra cost is a repair cycle. The 15 builds that finished unchanged scored about seven functional points higher than the 45 that needed edits, so a model that checks its work and changes nothing is itself a mild sign of a better build.

Cost is not only higher with a larger verification surface, it is also less predictable as shown in Table VI. To measure that unpredictability i.e. how much cost varies between runs whose setup is identical, we use the repeat runs. Every model built every application under every core condition four or five times, and those repeats are identical in everything we control, the same prompt, the same model, and the same tools. Any difference in what they cost is therefore run-to-run variation rather than the result of changing something.

The calculation has two stages. In the first stage, we take a single combination of one model, one application, and one condition, list the token cost of each of its four or five repeat runs, and compute the standard deviation of those costs, which measures how far the individual runs sat from their own average. That produces one number for that combination, its cost spread. In the second stage we take those numbers across all the combinations belonging to a condition, 42 of them for each core condition since six models each built seven applications, and report their median.

The medians are shown in Table VI with boot_check the most predictable at 47K tokens, with delayed_verification the widest in the study at 408K tokens. The boot probe is therefore the most predictable configuration as well as the cheapest. Granting a shell (execution) roughly doubles the spread against building blind, adding screenshots widens it further, sight without a shell is wider still (301k), and the delayed verification workflow is the least predictable of all. For anyone budgeting agent runs, this matters alongside the median cost, because a surface that looks affordable on average is still hard to plan around if any individual run may cost far more.

Figs. 4 and 5 draw the cost-quality frontier (each condition plotted by what it costs against the quality it delivers) twice: once against the machine-probed behavioral score (Fig. 4) and once against the human-graded interface score (Fig. 5). The boot_check’s efficiency and the shell's cost premium hold in both views. We keep the two quality judgments separate on purpose, for three reasons. First, they measure different things. A probe checks whether behavior is correct over the API, while the human judges whether a person can actually use the application, so a composite would average measures whose meaning differs across tasks. Second, any weighting of the two into a single number would be arbitrary, and a different weighting could reorder the conditions. Third, keeping them separate lets a reader check that the frontier's shape does not depend on which judge you trust, which is itself a finding.

Both figures pool the six models (the opening of Section IV explains why pooling is consistent with the design). Fig. 6 draws the same frontier separately for each model, where the same shape recurs at each model's own quality level.

TABLE VI
MEDIAN COST SPREAD BY CONDITION

| Condition | Tasks and Model combinations | Median run-to-run spread (token count) |
|---|---|---|
| no_verification | 42 (7 × 6) | 81k |
| static | 42 (7 × 6) | 77k |
| boot_check | 42 (7 × 6) | 47k |
| execution | 42 (7 × 6) | 149k |
| visual | 42 (7 × 6) | 188k |
| delayed_verification | 12 (2 × 6) | 408k |
| visual_no_shell | 18 (3 × 6) | 301k |
| behavioral | 6 (1 × 6) | 91k |

Medians in the sealed tables take the upper of the two middle values; medians quoted in the running text average the two. The conventions differ visibly only for the behavioral arm's total (375k vs 369k).

### D. *The Sight Boundary*

We measured whether screenshot access (visual) improves visually graded work beyond what a shell (execution) already delivers. The averages favor sight everywhere it should help, but the effect does not survive holm correction. How much a model looked at its screenshots did not line up with how much it gained.

We compared mean human interface scores over all assigned runs (Table VII), with non-starters counted as zero, and conditions with screenshots, visual and visual_no_shell, score higher than execution, by +10.9 and +4.3 on the calendar-week-view for visual and visual_no_shell respectively, and by +2.9 and +5.7 on the metrics-dashboard.

But our pre-specified test is not convinced that the advantage is real. That test, P4, pools only the two visual-versus-execution comparisons, and gives +6.9 points, 95 percent confidence interval [+0.8, +13.4], permutation p = .041, which does not survive the six-way family holm-corrected p (corrected p = .083).

So, sight's benefit on visually graded work is suggestive, not established, in this batch. The secondary contrast that separates seeing from running i.e. visual_no_shell versus execution (S2 in Table III) points the same way with the same caution (visual_no_shell versus execution: +5.0, interval [−1.4, +11.8], Fig. 7).

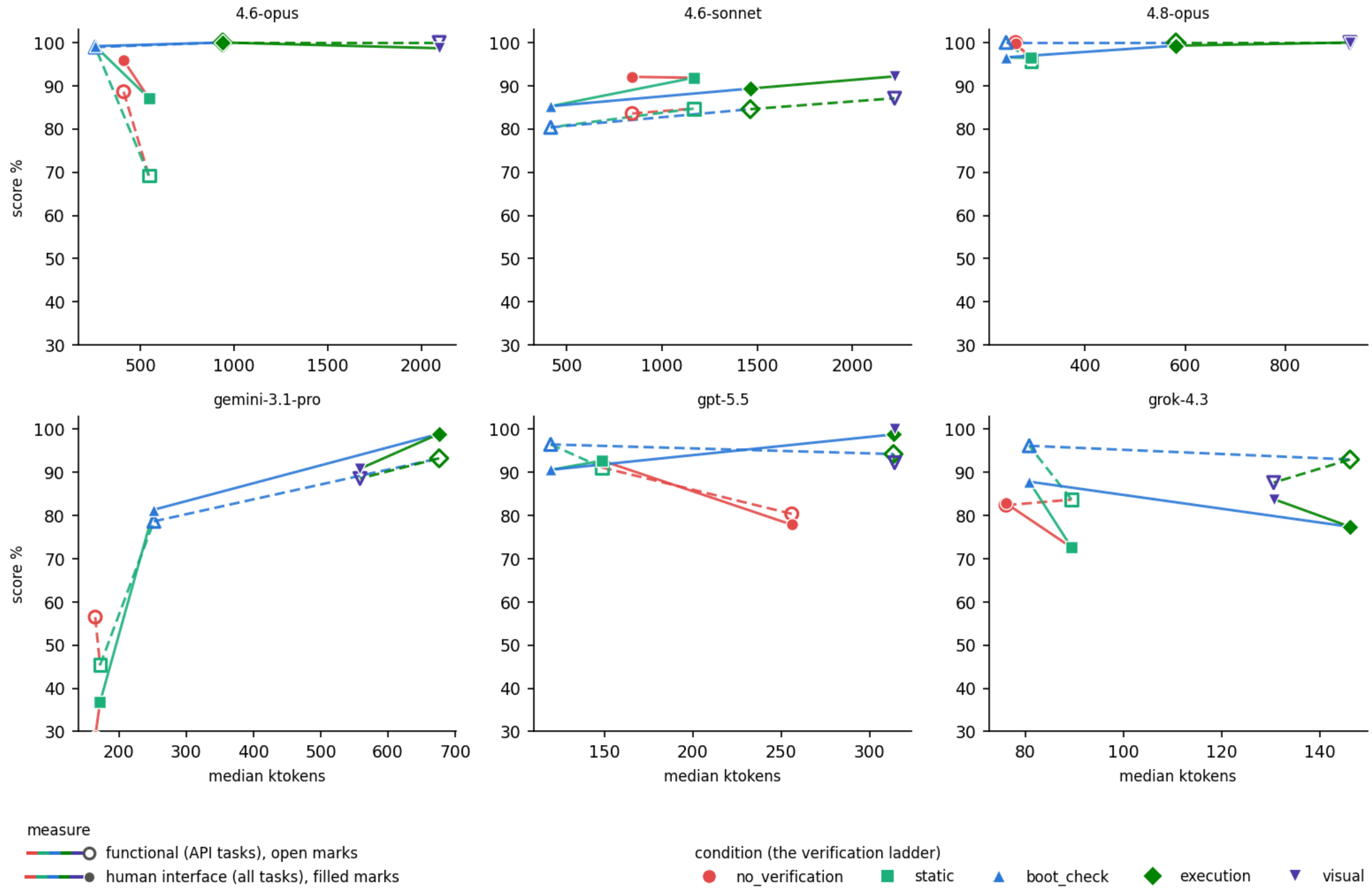

FIGURE 6. The same frontier as Figures 4 and 5, drawn separately for each model. In every panel x axis is that model's median token cost, marker fill gives the measure (open marks = functional score on the API tasks, filled marks = human interface score on all tasks), and marker shape and color together give the condition. The markers are connected in ladder order rather than cost order, so a leftward step means a rung that costs less than the one below it. Each segment carries the color of the rung it leaves. The shape recurs at each model's own quality level, which is why pooling in Figures 4 and 5 is representative.

TABLE VII
HUMAN INTERFACE SCORES

| Condition | calendar-week-view | Dashboard |
|---|---|---|
| no_verification | 81.4 | 64.3 |
| static | 84.8 | 60.4 |
| boot_check | 96.6 | 82.5 |
| execution | 86.4 | 92.0 |
| visual | 97.3 | 94.9 |
| visual_no_shell | 90.7 | 97.7 |

TABLE VIII
VISUAL VERSUS EXECUTION PER MODEL

| Model | calendar-week-view gap (no. of screenshots) | metrics-dashboard gap (no. of screenshots) |
|---|---|---|
| claude-4.6-opus | -0.9 (36) | 0.0 (48) |
| claude-4.6-sonnet | +23.2 (62) | -2.9 (51) |
| claude-4.8-opus | 0.0 (9) | +1.0 (21) |
| gemini-3.1-pro | -0.9 (34) | -19.2 (11) |
| gpt-5.5 | +0.9 (3) | +2.9 (22) |
| grok-4.3 | +42.9 (1) | +35.6 (0)[a] |

[a]Non-user: this model took no screenshots in these runs. Under the pre-specified rule for the per-model split, a non-user's sight effect is untestable. The value is shown for completeness. Because the model never looked, the gap cannot come from seeing the page and it reflects the model's unusually low score with the shell alone (Section IV-D)

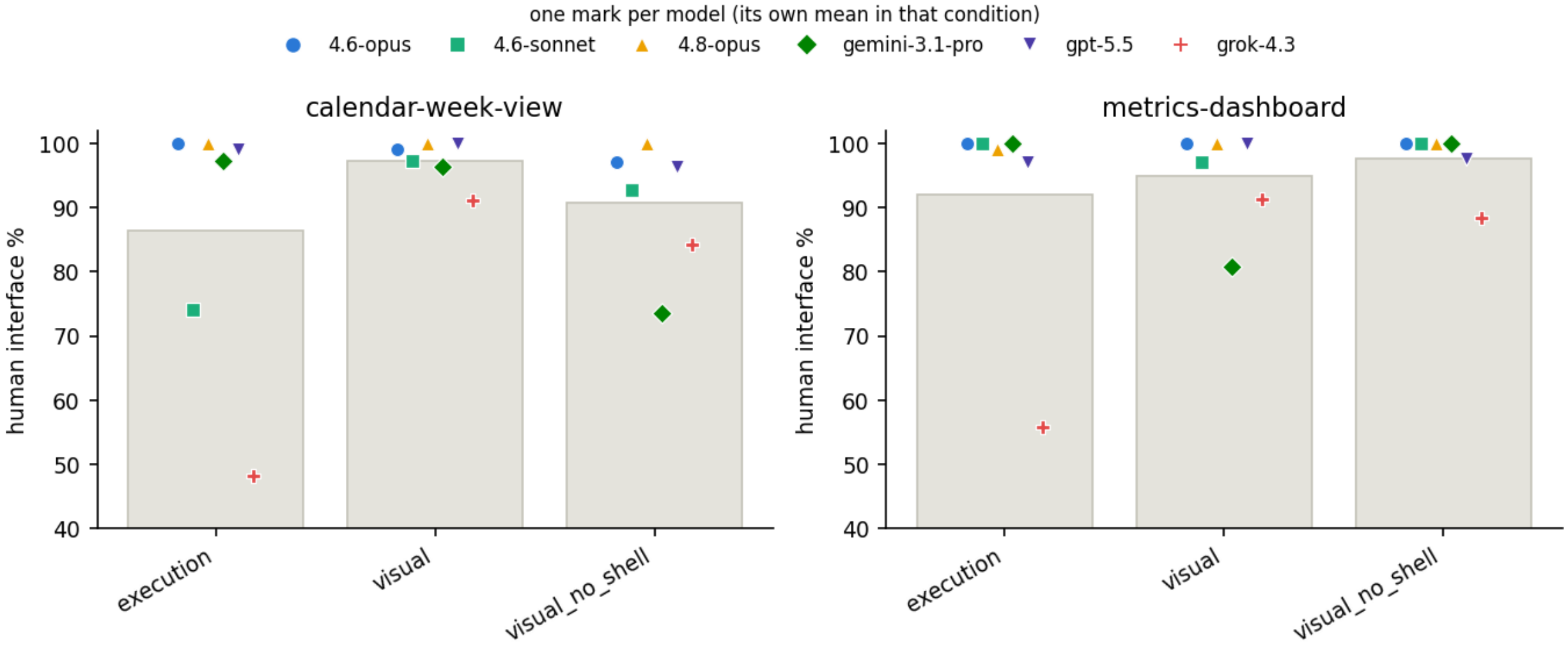

FIGURE 7. Sight-without-shell matches sight-with-shell on both visual venues, locating whatever value exists in the visual channel itself rather than the shell beneath it. Each bar is the mean human interface score across all six models in that condition, and each mark on it is one model's own mean, colored and shaped per model and spread horizontally so that models landing on the same score stay countable. The same model keeps the same mark in both panels, so a model can be followed across conditions and tasks. Both panels carry the same axis, repeated on the right for reading convenience.

We would like to mention two patterns that hold throughout the averages, but they are descriptive and we report them without a statistical test. First, how much sight helps on a task depends on whether that task's typical mistakes are visible in a picture (Table VII). The calendar-week-view's mistakes are the most visible kind, misplaced events and interactions that do not work, and sight helps most there (+10.9). The metrics-dashboard's mistakes are static layout, which a model can partly get right from the code alone, and sight helps less (+2.9). The log explorer's mistakes are lag and stutter, which no still image shows, and there the visual condition lands slightly below the shell (-2.2, Section IV-F). Second, removing the shell did not remove sight's value. On the metrics-dashboard, the condition with screenshots and no shell (visual_no_shell) scored as well as the full visual condition, in fact slightly better (97.7 versus 94.9). This suggests that whatever sight adds, comes from the seeing itself and not from the shell underneath.

However, the per-model split complicates attribution. Table VIII restricts the visual-versus-shell contrast to individual models. The pattern is awkward for a simple "sight helps" conclusion, because use and benefit do not line up. The largest gap belongs to grok-4.3, which gained 42.9 points on the calendar-week-view task while taking a single screenshot, and a further 35.6 points on the dashboard while taking none. The pooled average is driven largely by that model. Most of the other models were already near the maximum with a shell alone, so their gaps are small, with one exception, claude-4.6-sonnet, which took the most screenshots on the calendar-week-view (62) from a lower shell baseline and gained 23 points there. No other per-model contrast is individually significant. How much a model looked did not track how much it gained, so we cannot credit the pooled benefit to looking.

Fig. 8 shows the problem in one picture. Each point is one model on one visually hard task, placed horizontally by how many screenshots it took and vertically by how much better it scored with screenshots than with a shell. If looking caused the improvement, the points should climb from left to right, heavy users high and non-users near zero. They do not. The highest points sit at the far left with the model that barely looked, one of the heaviest users also sits high at the far right, and the models between are near zero. What does line up with the gains is the shell starting point. The two biggest gainers, grok-4.3 and claude-4.6-sonnet, scored low with a shell alone (around 48 to 56 for grok, 74 for sonnet on the calendar) and so had the most room to improve; the models already near the top barely moved however many screenshots they took. One moderate user, gemini-3.1-pro, took 11 screenshots on the dashboard and scored 19.2 points worse there. For grok, which barely used the tool, two explanations fit our data and we cannot tell them apart. grok's shell builds were unusually bad on their own (they shipped the study's worst interaction defects, Section V), or merely being granted a screenshot tool changed how it wrote its code even unused. Separating these would take a future experiment.

What this study can support is therefore modest. Granting sight never hurt the visually graded tasks on average, and its pooled benefit is positive but does not survive holm correction. Because the gains landed on the models with the most room to improve rather than on the models that looked the most, we cannot credit them to models actually looking at their pages and fixing what they saw.

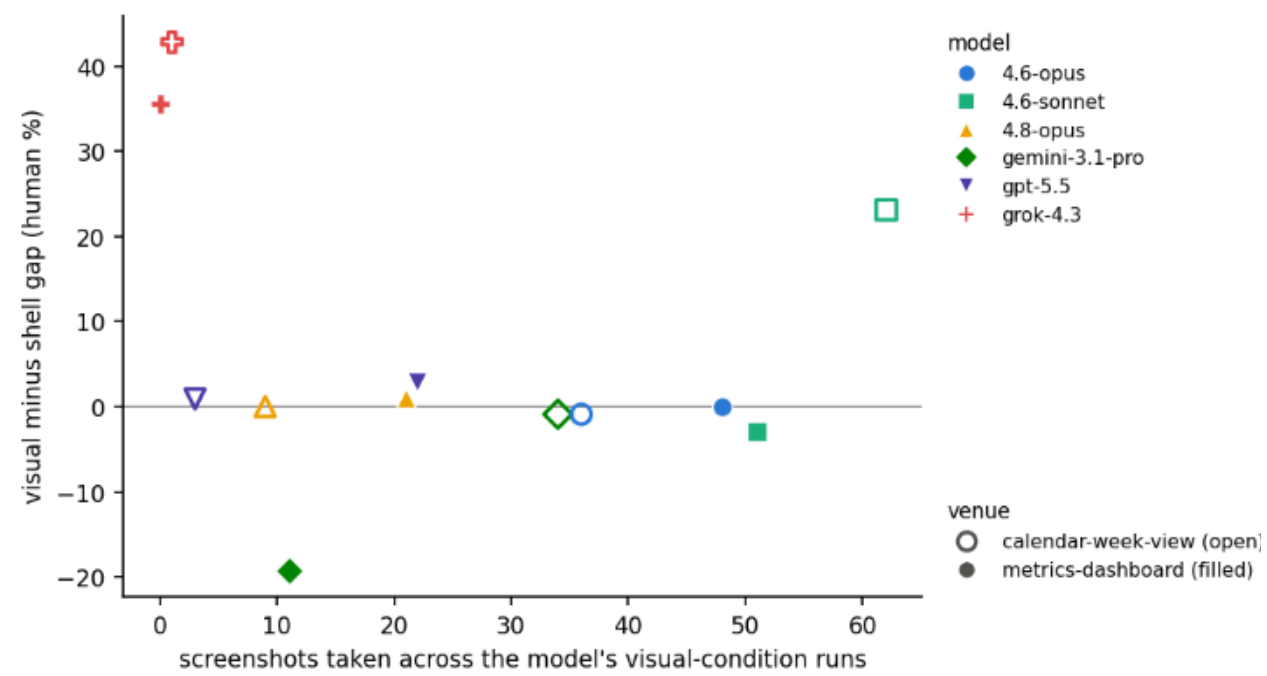

FIGURE 8. Each point is one model on one visual task, keyed by the same color and shape the model carries in Figure 7, with the task given by fill (open = calendar, filled = dashboard). Horizontal position counts the screenshots it took and vertical position shows how much better it scored with screenshots than with the shell. If looking caused the improvement, the right side should sit high. Instead, the largest gain (top left) belongs to the model that took one screenshot, and most heavy users sit near zero, with one high exception at the far right.

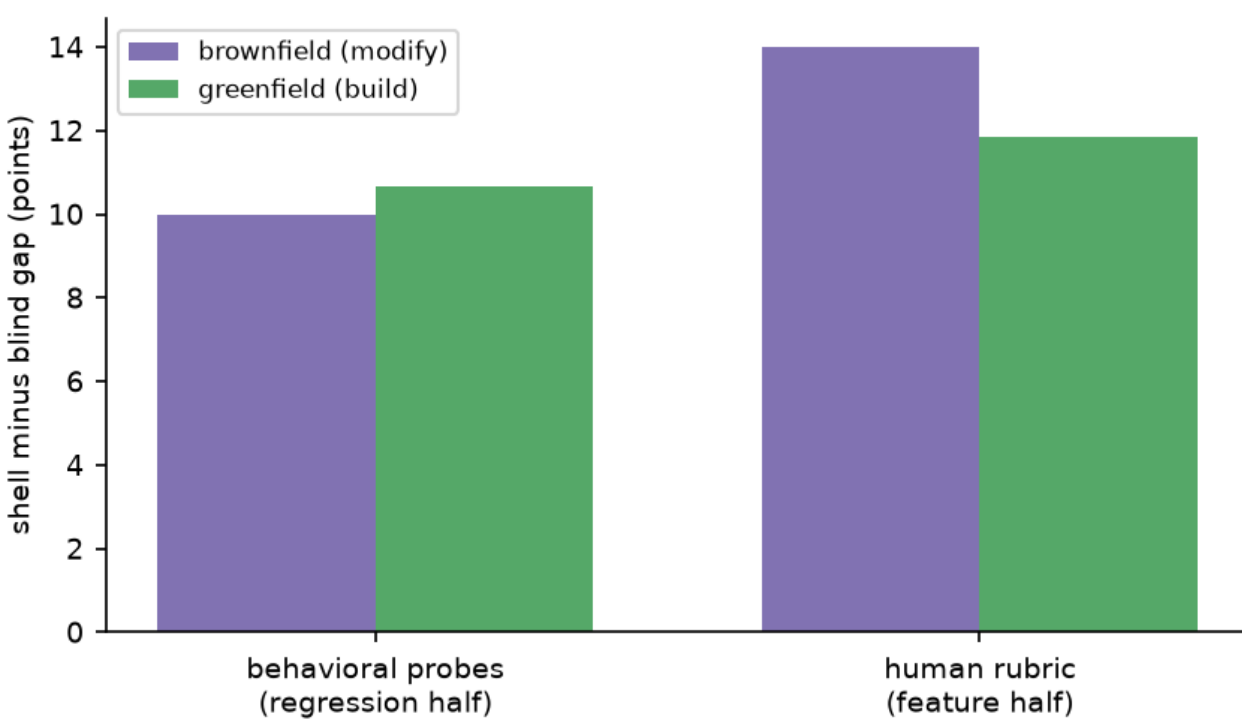

FIGURE 9. The shell-versus-blind gap on the modification task (brownfield) next to the same gap on the greenfield build of the same board, for both graded halves. The bars match within about a point: modifying unfamiliar code showed the same tool sensitivity as writing fresh code, the pre-specified null.

### E. *Brownfield modification: a clean null*

We tested whether verification tools matter more when an agent modifies an unfamiliar codebase than when it writes fresh code. We found in our study that they matter about the same amount.

The modification task is graded in two halves, a regression half asking whether everything the original board did still works (its functional probes rerun verbatim), and a feature half asking whether the newly requested labels are done well (a human-graded rubric). On the modification task the execution versus no_verification gap is 10.0 points on the regression half and 14.0 points on the feature half. On the greenfield version of the same board the matching gaps are 10.6 and 11.8 (Fig. 9). The two functional gaps are almost identical (10.0 versus 10.6) and the two human gaps are close (14.0 versus 11.8), so modifying someone else's code showed about the same tool sensitivity as writing fresh code, not more.

The pre-specified test asks the question directly. We measured the tool gap (execution versus no_verification) twice for functional score, once on the modification task and once on the fresh build of the same board, then subtracted one gap from the other, a design called a difference-in-differences. The subtraction comes to -0.7 points, essentially zero, which means it matters about the same in both settings. The 95 percent confidence interval around it runs from -13.7 to +12.7, which is wide.

In other words, the data cannot say which way this goes. It could be slightly less, slightly more, or the same. What the interval does rule out is a large difference in either direction. Had modifying existing code made the tools matter by roughly thirteen points or more, a study of this size would have caught it. A smaller difference could still hide in the noise, so a honest reading is that we found no difference, which is not the same as showing that none exists.

We report this as a null with one caveat. Our brownfield starter is small (a few files), well structured, and freshly seeded, which is the friendliest possible setting for modification, so the hypothesis could still hold on larger or messier codebases.

We conducted an additional check that asks whether the agents actually modified the code we gave them or quietly threw it away and rebuilt from scratch, which would make the comparison meaningless. We tested this mechanically. Each submitted file was compared against the six starter files it began from using content fingerprints, so any changed byte counts as a change, and a build was classed as a rewrite if it changed or deleted more than half of the starter. The result supports the comparison. 140 of 144 builds kept the starter and extended it, the median build touched exactly three of the six starter files in every condition, and the four rewrites are spread across different conditions (three in static, one in visual) rather than piling up where the tools would have made a rewrite easy. The rewrites look like isolated flukes, not something the tools caused.

### F. *Performance budgets: measurement beats sight*

We measured the same tool ladder on the log-explorer-perf, the one task whose failures can only be measured, not seen. The finding is a reversal. Section IV-D showed that on the two visually hard tasks (metrics-dashboard and calendar-week-view) the visual condition sat at or above the execution condition on the human interface score, 97.3 against 86.4 on the calendar and 94.9 against 92.0 on the dashboard (Table VII). On the log-explorer-perf the same comparison turns around. The execution condition moves to the top at 93.8 and the visual condition falls behind at 91.7, because a screenshot is a still image and these failures live in motion and delay. The reversal lands exactly where the study was built to expect it.

This was a designed prediction rather than an unexpected outcome. We put the log-explorer-perf in the study for exactly this purpose, as the task where screenshots should be useless and running the app and interacting with it should be the most effective way to test it. The log-explorer-perf's failures do not show up in a screenshot and cannot be caught over the API.

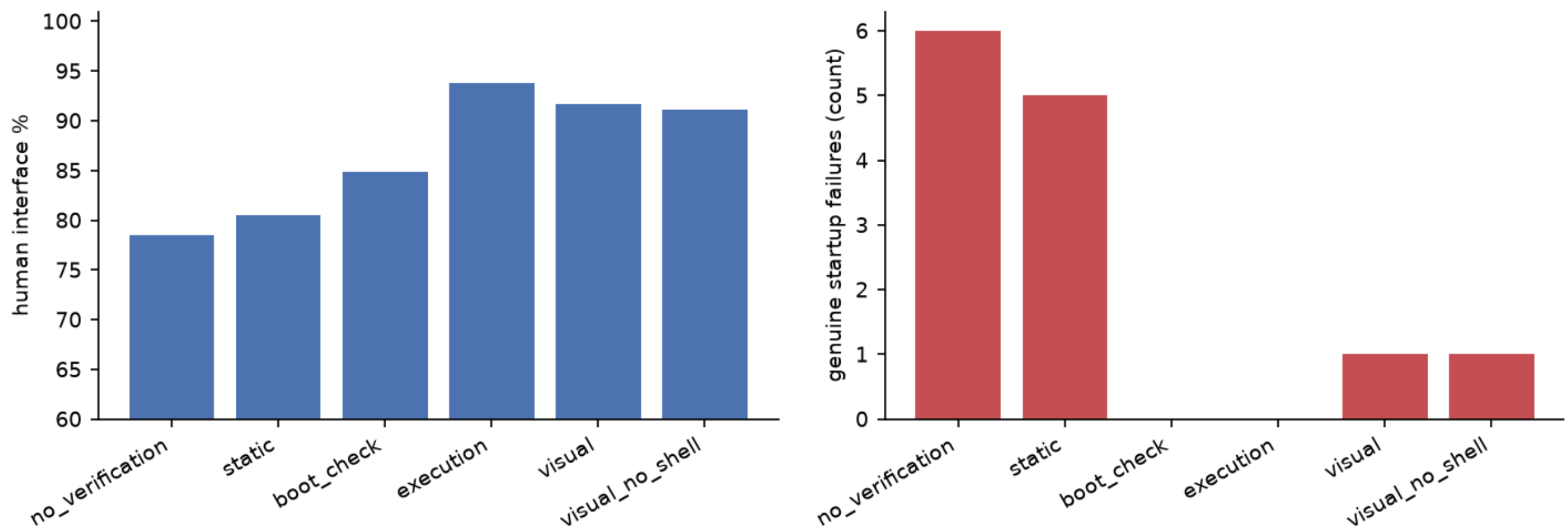

FIGURE 10. The log explorer reverses the visual tasks' ordering. Left: human interface scores by condition. The shell leads and both screenshot arms sit slightly below it, because a static image cannot show jank. Right: this task's 13 genuine launch failures by condition, nearly all in the blind and linter arms. Much of the shell's headline advantage here is startup survival, not rendering quality.

We can only see them by running the app with a lot of data, where a frozen scroll or an unvirtualized list finally appears. So the negative control behaved as designed. The committed contrast (S3 in Table III) finds that adding screenshots on top of the shell buys nothing on this task (−2.2 against the shell, interval [−12.5, +7.3]), and the screenshot-only arm tells the same story descriptively. visual_no_shell averages 91.2 against execution's 93.8 (Fig. 10). The control passes both ways, and it confirms the rule this study keeps finding, a checking tool helps only when it can observe the kind of failure the task actually produces.

This task also produced 13 of the study's 66 genuine launch failures, and nearly all of them (11 of 13) fall in the no_verification and static conditions (Fig. 10). That fits how much setup the task needs. Before the app can show anything, it has to load a database full of log data and run the code that queries and serves it, so there are more moving parts that can break before the first page ever loads. The execution versus no_verification effect here is +15.3 points, interval [+7.3, +24.0], raw p = .023, which does not survive the family correction (corrected p = .068).

That effect is also uneven across models in a revealing way (Table IX). Almost all of the +15.3 points come from one model, gemini-3.1-pro, whose no_verification builds all failed to even start and scored zero while its execution builds ran fine, giving it a full 100-point gap.

So on this task, the execution's apparent quality edge is mostly just its builds starting at all, dressed up to look like quality. One limit on the scope of verification is worth mentioning. The spec sets exact speed targets, but we do not yet have a tool that measures them, so the human grader judged the visible signs of poor performance instead, such as freezes, blank regions, and an overgrown page. This section therefore reports what the grader saw, not timed measurements.

### G. *Self-authored tests are not a substitute*

We measured whether a model that can only run its own test suite (the behavioral condition) reaches the quality of a model given the full execution shell.

TABLE IX
LOG EXPLORER MEAN HUMAN INTERFACE SCORE: EXECUTION VERSUS NO_VERIFICATION, BY MODEL

| Model | no_verification | execution | Gap | Launched (no_verification) | Launched (execution) |
|---|---|---|---|---|---|
| claude-4.6-opus | 100.0 | 100.0 | +0.0 | 5/5 | 5/5 |
| claude-4.6-sonnet | 100.0 | 95.0 | -5.0 | 5/5 | 5/5 |
| claude-4.8-opus | 100.0 | 100.0 | +0.0 | 5/5 | 5/5 |
| gemini-3.1-pro | 0.0 | 100.0 | +100.0 | 0/5 | 5/5 |
| gpt-5.5 | 80.0 | 97.0 | +17.0 | 4/5 | 5/5 |
| grok-4.3 | 91.0 | 71.0 | -20.0 | 5/5 | 5/5 |

We found that it does not, and the gap is not only about quality of tests written. The shortfall has two prominent sources. Part of it is survival and fewer behavioral builds ever started, and no test suite can rescue an application that does not launch. The rest is a blind spot no well-written test can compensate for, which is a model's own tests encode its own understanding of the task, so when that understanding is wrong, the tests pass and the misunderstanding ships anyway.

The behavioral condition runs on seat-booking only, 24 runs, and it is the weakest configuration in the study. Its survival is 83 percent and its functional score is 71.2, against 87.3 for execution on the same task and 94.2 for execution pooled across all API tasks (Fig. 11). Breaking the score down, the shortfall sits in two places as mentioned earlier. Fewer

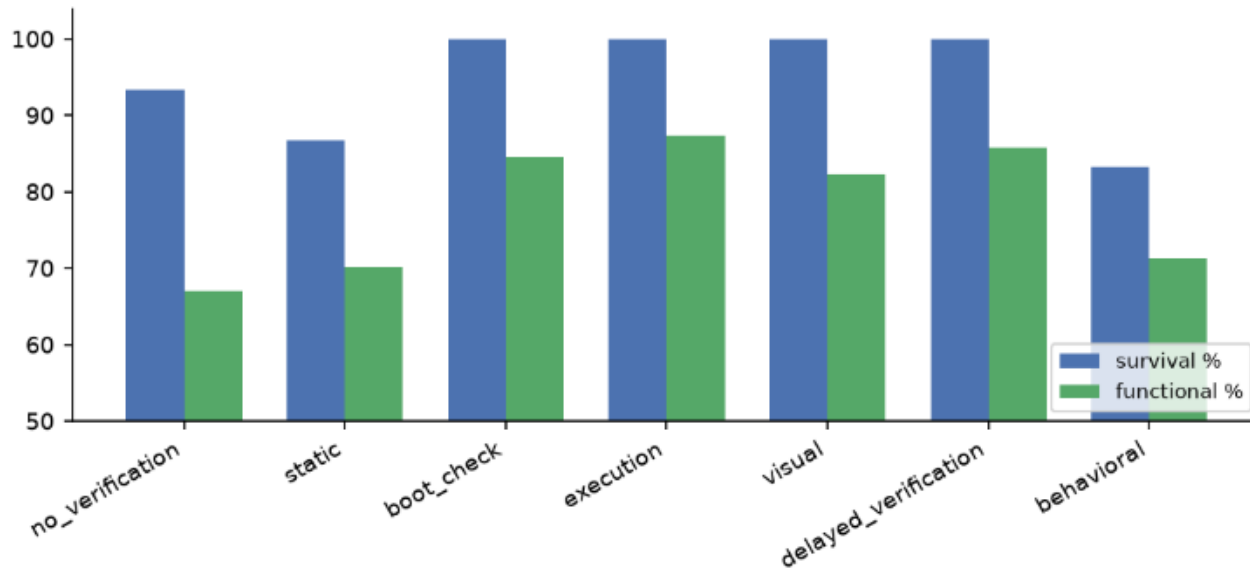

FIGURE 11. The one task that ran the self-authored-tests arm. That arm (rightmost) is the weakest surface on both survival and functional score: under a fixed budget, writing a test harness appears to displace finishing the application.

builds survive at all, and the ones that do run struggle on the deep concurrency probes.

The behavioral arm did not stop because it hit any limit. No run was cut short by the harness or a token budget, every run ended by declaring itself finished, and these runs stopped far below what the shell runs spent. On seat-booking it used 369 thousand tokens against the shell's 810 thousand and stopped after 22 steps against 49, while writing a similar amount of code. It wrote its tests, watched them pass, and finished early. A model's own tests only check whether its code matches its own understanding of the task. If the model's understanding is wrong, the tests could be wrong in the same way and still pass. A shell is different, because running the application shows what actually breaks.

The pre-specified secondary contrast (S5 in Table III: behavioral minus execution on the functional score) is −16.1 points, 95 percent interval [−31.7, −1.7], p = .043, and it is negative for five of the six models. So, the models that had to write their own tests scored about 16 points lower on the automated behavior probes than the models that were provided with a shell. The whole interval sits below zero, so the shortfall is unlikely to be chance, and it holds for five of the six models. The interval is wide, running from about 2 points to about 32, so we are far more confident about the direction of the gap than about its size. Though this is a small sample on a single task, so we report it as a secondary finding, not a headline result.

## V. The Defect Catalog: How Applications Actually Fail

These examples show the typical ways applications failed, not every case we found. The human grader's full notes record many more incidents than the six below, which were chosen because each defeats a different verification layer. The point of the catalog is not a census of everything that went wrong but a demonstration that consequential failure classes exist that no unit test, no linter, no boot probe, and no API probe would catch. The pattern that unifies them is that functionality was verified at one layer but broken at another.

- *The inert page*: This is a calendar-week-view defect where the API is perfect and rendering is pixel-accurate, but a transparent time label layer stretches over the whole grid that does not pass the click events through, so every mouse interaction dies silently. The model, running in the execution condition, had even written automated tests for the code that calculates where each element sits on the grid, the coordinates and sizes that place the time labels and cells. Those tests passed, so the positioning math was provably correct. But a unit test only checks that logic on its own. It cannot notice that an invisible layer is sitting on top of the grid and swallowing every click, which is what made the page unusable. No API probe, no code review, and no passive screenshot can see this. Only clicking can.
- *The server that cannot serve itself*: In development mode, the frontend forwards some web addresses to the backend so the page can reach its data (a proxy rule). Two models independently wrote that rule so broadly that it also captured requests for the application's own JavaScript module. The browser asked for the page's program files and was handed data instead, so the page loads without its code. Both builds had been verified only through their production path, where the proxy is not involved.
- *The date that became a number*: In one model's no_verification run on the calendar-week-view, the layout code clamps each event with a numeric max function applied straight to Date objects, Math.max(start, dayStart), which silently collapses them into millisecond numbers the renderer can no longer position. The data was fully correct and almost nothing drew (scored 7.1 percent, with none of its eight rubric items fully passing). The same clamping mistake appears again in the static condition, though the replicate there re-wraps the value at each use and still renders, so it is not a clean failure. The same model's sight-granted runs wrap the clamp back into a date, new Date(Math.max(..).), and render correctly.
- *The linted syntax error*: In the static condition, one build ships an unparseable frontend file. The logs show the model invoked the linter exactly once, but it had never added a lint script, so the call failed with "Missing script: lint" and inspected nothing. The model ignored that failure, then a later edit spliced a fresh syntax error into the file (an old_str that ended mid-identifier, leaving a dangling fragment), which was never re-checked before finishing. Granting a tool is not using it, invoking it is not wiring it up, and neither is the same as heeding it.
- *The immortal error banner*: A dashboard whose stylesheet overrides the HTML mechanism its own JavaScript uses to hide the "backend unavailable" banner, so a healthy application permanently tells the user it is down.
- *The query that works until asked*: One build stores its dates as plain text in the database, and one of its queries does arithmetic on those dates. The server starts cleanly, so a boot probe passes it. The crash comes on the first real request, when the arithmetic meets the text, a failure that by construction exists only at request time.

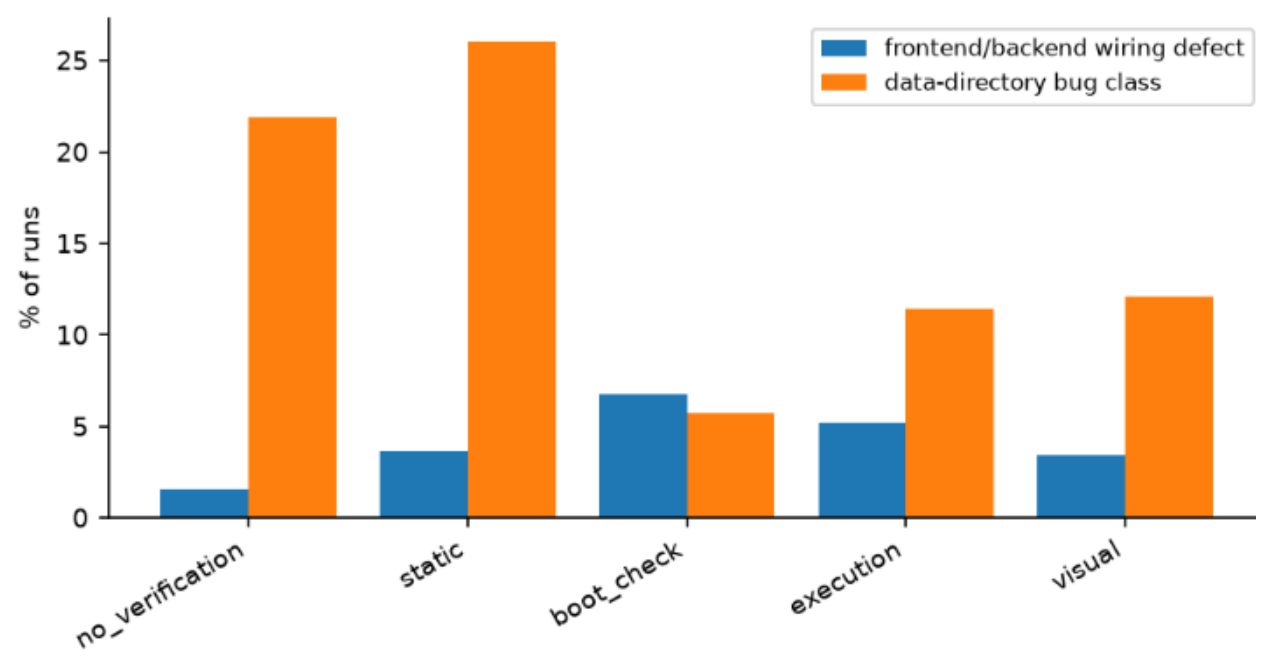

FIGURE 12. Configuration-incoherence rates by condition. Running the dev script (execution condition) reduces but does not eliminate the class. The boot probe nearly eliminates the data-directory bug because its one call exercises exactly the failing step.

Two of these failures are not one-offs. The same Vite misconfiguration (Vite is the standard frontend development server these applications use) recurs in 52 runs, splitting into a proxy that swallows the app's own api.js (20 runs, almost all claude-4.8-opus) and a Vite launched from the wrong folder so it ignores the proxy config (32 runs, almost all claude-4.6-opus), which suggest this is a model behavior rather than random noise.

These counts come from the human grader's hazard log, which records every run where the grader encountered the defect while grading. Section VI-A's scanner counts (45 wiring defects; 22 proxy-not-loaded; 16 api.js collisions) come from a static scan applied uniformly to all 1,116 builds. The two instruments measure related but not identical conditions, so their totals differ by design; both datasets are released (grading-hazards.csv, wiring_report.csv).

## VI. Configuration Integrity and Model Signatures

### A. Declared-launch integrity

The launch-configuration defects introduced in Section V were found and then measured in three steps. First, while hand-grading, the human grader kept running into the same launch problems and wrote each pattern down as it recurred. Second, once the patterns were clear, we turned them into two automatic scanners and ran both over all 1,116 builds. The scanners read the code without running anything, so every build is checked the same way no matter its tool condition or whether it even starts:

- *Wiring scanner*: does the frontend's declared start-up actually load its own proxy configuration, and does that proxy point to the port where the backend listens?
- *Data-directory scanner*: does the code create its database directory before trying to use it? This is the defect behind several startup crashes.
- Third, the few builds the scanner could not classify on its own were classified by hand, and those classifications are recorded in the released dataset (datadir-unknown-resolutions.csv). No build was modified at any point.

We single out the first two defect classes, out of everything in the grading notes, for two reasons. They recurred most often during hand-grading, and both can be checked mechanically and identically for every build.

45 of 1,116 builds (4 percent) ship a wiring problem. That rate is low and fairly even across conditions, roughly 1.6 to 6.8 percent, and is highest in the boot_check condition. The data-directory bug is much more common, 169 builds (15 percent), and it depends heavily on tools, about 21 to 26 percent in the no_verification and static conditions, dropping to 5.7 percent under the boot probe and around 11 to 13 percent in the shell conditions (Fig. 12).

These defect rates carry two lessons. First, running the dev path helps but does not fix the problem. Several builds in the execution condition only ever checked their production build and shipped a broken development configuration anyway. Second, which model built the app predicts the defect better

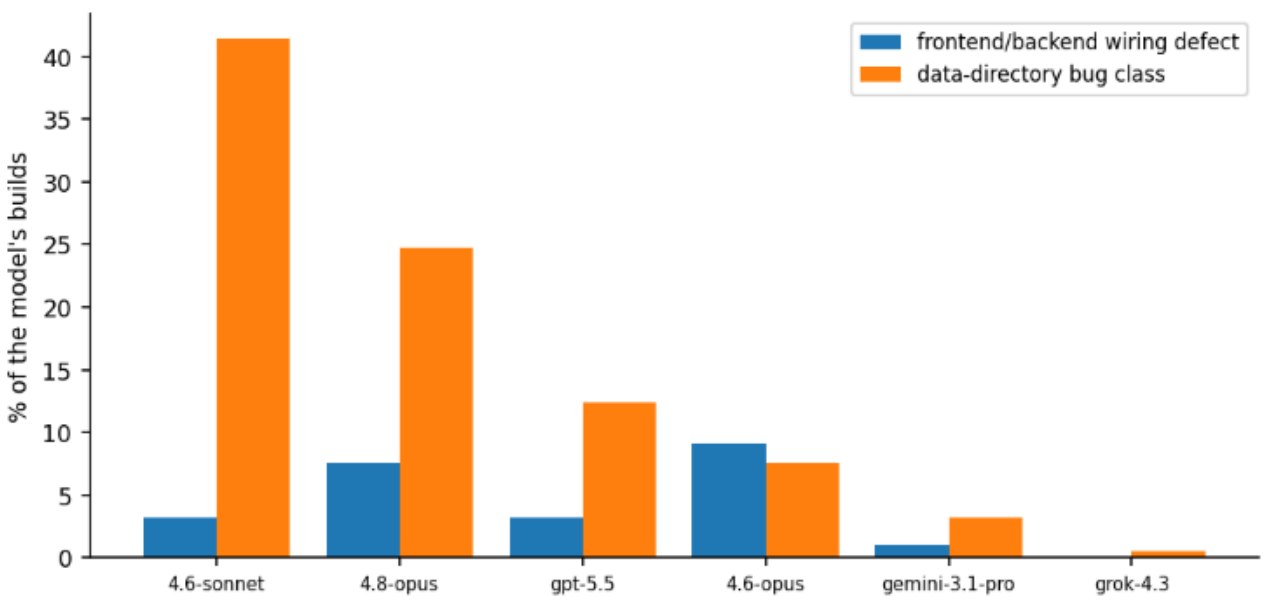

FIGURE 13 The same two defect classes split by model instead of condition. The spread is far wider than Figure 12's: which defect a build carries is predicted better by who built it than by what tools it had, the temperature-zero idiom finding in quantitative form.

### B. Model signatures (exploratory, observational)

We ran a batch-wide structural scan to determine whether the implementations carry a stable model signature independent of its tool condition. The structural scan extracted ten concrete engineering choices from every artifact: where the package manifests live, how the directory tree is organized, which JavaScript module system is used, how the dev processes are orchestrated, how the frontend build tool is invoked, how the frontend calls the API, whether TypeScript is used, whether the standard build tool is used, whether tests are written, and whether a README is written. For each structural feature we computed its association strength (Cramér's V [35], a 0-to-1 measure of how strongly a categorical choice is predicted by a grouping) with model and, separately, with condition (Table X).

Of the ten structural choices we tracked, seven are model habits. They are predicted far better by which model wrote the app than by which tools it had, so each model's choices read like a fingerprint. Only one choice, whether the app ships tests, is tied to the tool condition instead, and it belongs to the behavioral condition. And that result comes from the study design rather than from the models. The behavioral condition's only tool runs the project's own test suite, so a model there has to write tests before it has anything to run. Everything else about how an app is built comes from the model, not the tools.

TABLE X
CRAMÉR'S V VALUES

| Feature | V with model | V with condition | Verdict |
|---|---|---|---|
| package manifest layout | 0.46 | 0.16 | Model idiom |
| directory tree style | 0.40 | 0.10 | Model idiom |
| module system | 0.36 | 0.10 | Model idiom |
| dev process orchestration | 0.39 | 0.13 | Model idiom |
| frontend build invocation | 0.27 | 0.12 | Model idiom |
| API client style | 0.34 | 0.12 | Model idiom |
| README written | 0.70 | 0.18 | Model idiom |
| tests written | 0.17 | 0.66 | Condition effect |
| TypeScript use | weak | weak | Neither |
| standard build tool use | weak | weak | Neither |

Each model's most common choices read like a fingerprint. One model, gpt-5.5, lays its project out as a src tree in 77 percent of its runs and starts its frontend build with a bare invocation in 94 percent, regardless of which tools it was given. These are style preferences rather than mistakes, and the bare invocation is in fact the variant that avoids the launch hazard described in Section VI-A. Another model, grok-4.3, mixes root and nested manifests in 82 percent of its runs. The launch-configuration hazards of Section VI-A are similarly concentrated, for example claude-4.6-opus accounts for 16 of the 22 cases where the frontend never loads its proxy configuration, and claude-4.8-opus for 14 of the 16 cases where a root-level api.js collides with the proxy rule. Code volume is essentially flat across the five core conditions (716 to 800 mean lines), so the shell's 2.35x token premium buys verification and iteration, not more software.

These model signatures are observational rather than experimental. We varied the tools on purpose, but the models are fixed groups within which we compare rather than randomized treatment across models, and the mix of runs per model is not perfectly matched, so we read them as patterns, not proven causes. Even so, they mechanically explain why defects cluster by model, since a defect usually grows out of one model's habitual way of building a project. A model makes the same design and appearance choices again and again, whether or not it is sampling deterministically, so those choices are properties of the model itself.

## VII. LIMITATIONS

### A. Passive Screenshot tool

Our screenshot tool takes a single argument, a URL. It renders a page and returns an image. It cannot click, type, drag, hover, or resize. This limits what the visual arms could ever have caught, and the run logs show the agents meeting that bound repeatedly rather than merely in principle.

Across the 264 runs in the two sight conditions, visual and visual_no_shell, 214 took at least one screenshot, making 1,326 calls to 260 distinct URLs. Of those calls, 512, or 39 percent, pointed somewhere other than the application's own root. These figures are a measure of effort spent reaching states the agent could not click its way into. In 110 runs the agent used at least one such URL, and in 67 runs it went further and authored a purpose-made page whose only role was to be photographed, with names such as preview.html, test-visual.html, viewport-360.html, and, in one seat-booking run, static-preview-hold.html and static-preview-booked.html, which are mock-ups of the states a click would have produced. The workarounds take three forms: Models forced state through the address bar, using fragments and query parameters such as #scroll-bottom, ?scrollTo=22, ?demo=edit, ?theme=dark, and ?width=360. They also built the harness pages described above. And in several runs they edited the application itself purely to make a state photographable, one agent noting that it would "temporarily make the API return dark unconditionally, screenshot, revert", and another adding "a GET variant of seed temporarily so I can hit it via screenshot navigation". One calendar run resorted to a javascript: URL that posted four events and reloaded the page. In 21 runs the model even stated the limit outright before working around it or giving up on the check. One wrote, "Let me verify the edit form opens by clicking. I can't click in screenshot tool. Let me trust the code." Some other models recorded that the tool "can't type into inputs", "can't resize", and "can't interact", each time falling back on reading the code instead of observing the behavior.

Two inferences can be made from this. Our visual conditions therefore measure the low end of what sight can do, since a tool that cannot click will never find a fault that only appears on clicking. And the inert page of Section V, a calendar that renders correctly and ignores every click, is precisely the defect this channel is structurally unable to see, which is why it survived the visual condition.

### B. Reasoning is not carried across turns

Because no reasoning content was returned to the agent, a model's reasoning could not be carried from one turn to the next, and every turn proceeded from the message history alone. This was identical across all conditions, so it does not distort the comparisons, but it does bound what the results speak to. Conditions differ in how many turns they take, and the arms with a shell run much longer than the arms building blind, so whatever benefit reasoning continuity might confer would accrue unevenly across the ladder. On an interface that returns and preserves reasoning, the quality gaps we report could widen and the cost gaps could narrow.

### C. Single Grader

The human interface scores come from a single grader. The grades were produced condition-blind, in shuffled order, against written per-item criteria. The repeatability of the scores was measured with a condition-blind regrade of a 10 percent sample after a four-week washout. The regrade process reproduced item verdicts at 98.8 percent (weighted kappa 0.97 pooled, 0.94 on launched runs) and agreed on every survival call (Section II-J). However, this addresses stability, not correctness. The limitation remains contained in scope and conditions such as whether an app started, what kind of crash it hit, and every automated behavior test are measured by code and return the same result on a rerun, so only the human quality ratings are exposed to this concern.

### D. Rubric-only tasks

Three tasks (the calendar, dashboard, and log explorer) have no automated functional tests, so the human rubric is their only grade. Grading one dashboard run exposed a weakness in that setup. An app can render its interface correctly yet be functionally empty, like a dashboard whose cards all display but hold no data. Because the rubric mainly checks whether the interface looks and behaves right, it can score such an app too generously and not penalize the underlying failure enough. We treat this as a known limit of the rubric's reach and flag it in the notes of each affected run rather than change the scores to correct for it.

### E. Statistical power

Two limitations of statistical power (a study’s ability to detect an effect that is really there) deserve plain statement. First, running the same model on the same task twice tends to produce similar code with the same quirks, so our four to five replicates of a cell are not four to five independent chances to observe a defect. If the model’s habit produces the bug, every replicate tends to share it. The study’s statistical strength therefore comes from having six different models, not from the replicates within each cell. Second, six models mean only six independent blocks. Comparisons pooled across the six models have reasonable power, but any comparison inside a single model rests on four or five runs per condition and cannot reach significance on its own, which is why per-model numbers in this paper are always descriptive.

### F. Correction and null results

Two of the six pre-specified effects look significant on their own but no longer clear the bar once we correct for testing all six together, so we report them as suggestive rather than established. The brownfield result is a null, but its confidence interval is wide, so it rules out only large effects and cannot exclude a small or moderate one. In short, where an effect survives the correction, this design supports it well. Where a result comes out null, all we can say is that any real effect is no larger than the far end of its interval. That is not the same as saying the effect is zero.

### G. Non-customizable sampling setting

Three of six models’ APIs rejected the sampling settings we requested and ran at the provider's defaults instead, but each such model used the same settings across all of its own runs, so there is no inconsistency within any single model.

## VIII. Conclusion

Across 1,116 controlled builds, the verification surface’s value separates into tiers that no single pass rate would reveal, and what decides each tier is reach: a tool changes the artifact only where it can exercise the part that fails.

The bottom tier, survival, is nearly free. Agents authoring blind fail to produce a launchable application in about one build in seven, almost always through environment assumptions and unforced source corruption, and one cheap, well-designed boot check removes almost all of these launch failures, at a fraction of a full shell's cost and with the most consistent results of any condition we tried. It is not quite perfect. One build in 192 still failed, and that one was a broken frontend on top of a healthy backend. The boot_check only watches whether the server starts, so a broken page is the single thing it cannot see.

The middle level of value is behavioral depth, getting the app to behave correctly under demanding runtime conditions. This comes only from actually running the code. The tests that expose it, handling many requests at once, surviving a restart with the data intact, and keeping the right order under load, are exactly what set the shell-equipped builds apart from every cheaper condition. Letting the model write its own tests instead is a poor replacement, and not because it runs short of room. On the one task where we tried it, the self-testing condition used less than half the shell's tokens and stopped after 22 steps against 49, while writing a similar amount of code. It wrote its tests, watched them pass, and finished early. Tests a model writes itself can only check its code against its own reading of the task, so they repeat a misunderstanding rather than catch it.

The highest level of value, how well the app renders and meets its measured targets, is where it matters most which kind of task you are on. On average, screenshots help exactly on tasks whose problems are visible to the eye, and add nothing on tasks whose problems can only be measured. The important caution is that even this average benefit does not pass the stricter corrected significance test, and most of it comes from a single model that barely looked at the screenshots at all. So two questions stay open: how large sight's real effect really is, and whether it works by the model actually looking at what it built. Our design can raise these questions, but this batch of data cannot settle them.

For practitioners the actionable summary is unchanged by the nuance:

- A cheap boot probe is the best verification token in this study.
- An unrestricted shell buys behavioral depth at 2.35 times the cost and leaves a characteristic residue of interaction dead frontends.
- Screenshots are cheap insurance on visually hard work but should not be mistaken for interaction testing.

Looking at each model on its own adds one more lesson. How much verification a model needs is itself a trait of the model. When we compare each model against its own no_verification baseline, the strongest model gained nothing from any tool, because it already scored perfectly on the API tasks while building blind. The weakest model gained 37 points once it was handed a shell. And the tool that helped each model most was not the same from one model to the next. So, the verification budget is best decided model by model, and for most models the cheap boot probe is enough.

For people who evaluate models, the study's method may matter more than its exact numbers. We graded the finished apps rather than the chat logs, kept every prompt byte-for-byte identical so that only the tools changed, decided in advance how to handle models that refused our settings and apps that failed to start, and kept a human judge for the one layer no automated check can reach. Together these choices exposed kinds of defects, and hard questions about what actually caused them, that a pass-or-fail benchmark is, by construction, unable to see. A single pass rate cannot distinguish never-started from almost-right, and it has no channel for a page that renders perfectly yet responds to nothing.

The single rule behind all of these results is reach. A boot probe watches startup, so it removes launch failures and nothing else. A shell drives the running application over its own interfaces, so it buys behavioral depth and leaves untouched the defects that live in the rendered page. Screenshots reach what is visible, so they pay on geometry and layout and add nothing where the fault is measurable but invisible. A linter reaches only the code it is actually wired to inspect, which in a quarter of its runs was nothing at all. Choosing a verification surface is therefore not a question of how much tooling to grant, but of whether the tools granted can exercise the way this particular application is likely to fail.

## Data and Code Availability

All data underlying the findings are openly archived at Zenodo under DOI 10.5281/zenodo.21961590 (the all-versions identifier, resolving to the latest release) and are browsable at https://github.com/achintmehta/coding-eval-agent-runs. The archive contains the 1,116 built applications exactly as shipped, complete per-run tool-call logs, all machine grades and condition-blind human rubric scores, the hand-audited launch-failure classifications, and three SHA256-sealed dataset snapshots, of which freeze-20260720 is the dataset of record that every number in this paper traces to. The regrade dataset can also be found under the Zenodo archive and github. The archive also contains the code that produced those numbers: the grading instrument and its frozen rubric cards, the verification chain that regenerates the analysis table byte-for-byte from the published inputs, and the pre-specified statistical plan with its committed analysis code and complete output. The agent harness that generated the builds is documented in the archive through its byte-exact prompts, per-condition tool schemas, one full example trace per condition, and every run's complete tool-call log. Appendix B maps each artifact class named in this paper to its exact path in the archive.

## Appendix A

### The Agent Interface

The two texts below are reproduced byte-exact (their hashes are verified against every run manifest), so their punctuation is the instrument's own.

Base prompt (constant across every run):

*You are a software engineer. Your workspace directory is {workspace}. Implement the task described by the user by creating files in the workspace using the provided tools. Work until the task is complete, then call finish. Use only the tools provided to you.*

Universal environment briefing (identical in every condition):

*Environment: the tools listed in this request are the complete set available to you. Tools not listed do not exist in this environment and cannot be installed or obtained. Review what is available and use whatever helps you complete and check your work. If no installation or execution tools are provided, author every required file completely from first principles; dependencies declared in package.json files are provisioned by the environment after you finish.*

Per-condition tool schemas (8 files) and one full example trace per condition are in the archive under agent-interface/ (Appendix B).

## Appendix B

### RELEASED ARTIFACT MAP

Everything this paper derives from is published in one archive: Zenodo DOI 10.5281/zenodo. 21961590 (all-versions identifier), browsable at https://github.com/achintmehta/coding-eval-agent-runs .

The dataset of record is the SHA256 sealed snapshot 20260720/, and the archive's README.md contains step-by-step reproduction recipes (regenerate the analysis table byte-for-byte, recompute every statistic, regrade any artifact from scratch).

**ACHINT MEHTA** received the B.E. degree in computer science India, in 2003. He has more than 20 years of software research and development experience spanning telecommunications, satellite communication, and cybersecurity domains. From 2003 to 2011, he worked in the telecommunications industry, developing protocol stacks, and working on GPON optical network terminal software for embedded real-time systems. For the last decade, he worked in the cybersecurity industry, first as a Senior Staff Engineer and Squad Leader developing intrusion prevention systems, cloud-based network security management, and zero trust secure access solutions, and currently as a Senior R&D Manager leading the development of several cloud native security solutions. His research interests include artificial intelligence, cloud security, zero trust architectures, network intrusion detection and prevention, telecommunication protocols, and cloud-native distributed systems.